\documentclass[twocolumn]{article}
\usepackage{geometry}

\usepackage{upgreek}
\usepackage{mathtools}
\usepackage{graphicx}
\usepackage{color}
\usepackage{float}
\usepackage{xspace}
\usepackage{booktabs} 
\usepackage{txfonts}
\usepackage{caption}
\usepackage{hyperref}

\newcommand{\apj}{Astrophys. J.} %{ApJ}    
\newcommand{\apjs}{Astrophys. J. Suppl. Ser.} %{ApJS}       
                                 
\newcommand{\araa}{Annu. Rev. Astron. Astrophys.} %{ARAA}       
\newcommand{\mnras}{Mon. Not. R. Astron. Soc.} %{MNRAS}     
\newcommand{\nat}{Nature}

\newcommand{\aap}{Astron. Astrophys.} %{A\&A}            
                                
\newcommand{\aapr}{Astron. \& Astrophys.\ Rev.} %{A\&AR}                                
\newcommand{\physrep}{Phys. Rep.} %{Physics Reports}     
\newcommand{\apjl}{Astrophys. J. Lett.} %{ApJL}          

\newcommand{\pasp}{Publ. Astron. Soc. Pac.} %{PASP}       
\newcommand{\pasa}{Pubs. Astron. Soc. Australia}

\usepackage[comma,super,sort&compress]{natbib}
\usepackage[noblocks]{authblk}

\newcommand{\rthree}{FRB~20180916B\xspace}
\newcommand{\frb}{FRB~20200120E\xspace}
\newcommand{\sgr}{SGR~1935$+$2154\xspace}
\newcommand{\gc}{[PR95]~30244\xspace}

\newcommand{\dmunit}{pc\,cm$^{-3}$}
\newcommand{\us}{$\upmu$s}

\graphicspath{{../}{figures/}}
\makeatother

\begin{document}
\begin{titlepage}
\newgeometry{top=1cm, bottom=2cm, left=2cm, right=2cm}
\onecolumn
\title{Burst timescales and luminosities link young pulsars and fast radio bursts}

%\include{auth_nimmo_ordered_initials}
%\correspondingauthor{K.~Nimmo}\email{k.nimmo@uva.nl}
%\include{auth}
% main
\author[1,2]{K.~Nimmo}
\author[2,1]{J.~W.~T.~Hessels}
\author[3]{F.~Kirsten}
\author[4]{A.~Keimpema}
\author[5]{J.~M.~Cordes}
\author[2]{M.~P.~Snelders}
\author[2]{D.~M.~Hewitt}
\author[6]{R.~Karuppusamy}
\author[7]{A.~M.~Archibald}
\author[8]{V.~Bezrukovs}
\author[9,10]{M.~Bhardwaj}
\author[1]{R.~Blaauw}
\author[11]{S.~T.~Buttaccio}
\author[12,13]{T.~Cassanelli}
\author[3]{J.~E.~Conway}
\author[14]{A.~Corongiu}
\author[15]{R.~Feiler}
\author[9,10,16,17]{E.~Fonseca}
\author[3]{O.~Forssén}
\author[15]{M.~Gawroński}
\author[18]{M.~Giroletti}
\author[19]{M.~A.~Kharinov}
\author[20,21]{C.~Leung}
\author[3]{M.~Lindqvist}
\author[18]{G.~Maccaferri}
\author[4]{B.~Marcote}
\author[20,21]{K.~W.~Masui}
\author[12,13]{R.~Mckinven}
\author[19]{A.~Melnikov}
\author[9,10]{D.~Michilli}
\author[19]{A.~Mikhailov}
\author[12]{C.~Ng}
\author[8]{A.~Orbidans}
\author[2]{O.~S.~Ould-Boukattine}
\author[4]{Z.~Paragi}
\author[9,10,22,23,24]{A.~B.~Pearlman}
\author[2,9,10,25]{E.~Petroff}
\author[26]{M.~Rahman}
\author[12]{P.~Scholz}
\author[20,21]{K.~Shin}
\author[27]{K.~M.~Smith}
\author[28]{I.~H.~Stairs}
\author[14]{G.~Surcis}
\author[29,30]{S.~P.~Tendulkar}
\author[3]{W.~Vlemmings}
\author[31]{N.~Wang}
\author[3]{J.~Yang}
\author[31]{J.~Yuan}

\affil[1]{ASTRON, Netherlands Institute for Radio Astronomy, Oude Hoogeveensedijk 4, 7991 PD Dwingeloo, The Netherlands.}
\affil[2]{Anton Pannekoek Institute for Astronomy, University of Amsterdam, Science Park 904, 1098 XH, Amsterdam, The Netherlands}
\affil[3]{Department of Space, Earth and Environment, Chalmers University of Technology, Onsala Space Observatory, 439 92, Onsala, Sweden}
\affil[4]{Joint Institute for VLBI ERIC, Oude Hoogeveensedijk 4, 7991 PD Dwingeloo, The Netherlands}
\affil[5]{Department of Astronomy and Cornell Center for Astrophysics and Planetary Science, Cornell University, Ithaca, New York, 14853, USA}
\affil[6]{Max Planck Institute for Radio Astronomy, Auf dem Heugel 69 53121 Bonn, Germany}
\affil{School of Mathematics, Statistics and Physics, Herschel Building, Newcastle University, Newcastle upon Tyne, NE1 7RU, UK}
\affil{Engineering Research Institute Ventspils International Radio Astronomy Centre (ERI VIRAC) of Ventspils University of Applied Sciences (VUAS), Inzenieru street 101, Ventspils, LV-3601, Latvia}
\affil[9]{Department of Physics, McGill University, 3600 rue University, Montréal, QC H3A 2T8, Canada}
\affil[10]{McGill Space Institute, McGill University, 3550 rue University, Montréal, QC H3A 2A7, Canada}
\affil[11]{Istituto Nazionale di Astrofisica, Istituto di Radioastronomia Radiotelescopio di Noto, C.da Renna Bassa Loc. Case di Mezzo C.P. 161, 96017,  Noto (SR), Italy}
\affil[12]{Dunlap Institute for Astronomy \& Astrophysics, University of Toronto, 50 St. George Street, Toronto, ON M5S 3H4, Canada}
\affil[13]{David A. Dunlap Department of Astronomy \& Astrophysics, University of Toronto, 50 St. George Street, Toronto, ON M5S 3H4, Canada}
\affil[14]{Istituto Nazionale di Astrofisica, Osservatorio Astronomico di Cagliari, Via della Scienza 5, I-09047, Selargius, Italy}
\affil[15]{Institute of Astronomy, Faculty of Physics, Astronomy and Informatics, Nicolaus Copernicus University, Grudziadzka 5, 87-100 Toru\'n, Poland}
\affil[16]{Department of Physics and Astronomy, West Virginia University, P.O. Box 6315, Morgantown, WV 26506, USA}
\affil[17]{Center for Gravitational Waves and Cosmology, West Virginia University, Chestnut Ridge Research Building, Morgantown, WV 26505, USA}
\affil[18]{Istituto Nazionale di Astrofisica, Istituto di Radioastronomia, via Gobetti 101, I-40129 Bologna, Italy}
\affil[19]{Institute of Applied Astronomy of the Russian Academy of Sciences, Kutuzova Embankment 10, St. Petersburg, 191187, Russia}
\affil[20]{MIT Kavli Institute for Astrophysics and Space Research, Massachusetts Institute of Technology, 77 Massachusetts Ave, Cambridge, MA 02139, USA}
\affil[21]{Department of Physics, Massachusetts Institute of Technology, 77 Massachusetts Ave, Cambridge, MA 02139, USA}
\affil[22]{Division of Physics, Mathematics, and Astronomy, California Institute of Technology, Pasadena, CA 91125, USA}
\affil[23]{McGill Space Institute~(MSI) Fellow}
\affil[24]{FRQNT Postdoctoral Fellow}
\affil[25]{Veni Fellow}
\affil[26]{Sidrat Research, PO Box 73527 RPO Wychwood, Toronto, ON M6C 4A7, Canada}
\affil[27]{Perimeter Institute for Theoretical Physics, 31 Caroline Street North, Waterloo, ON, N2L 2Y5, Canada}
\affil[28]{Dept. of Physics and Astronomy, University of British Columbia, 6224 Agricultural Road, Vancouver, BC V6T 1Z1}
\affil[29]{Department of Astronomy and Astrophysics, Tata Institute of Fundamental Research, Mumbai, 400005, India}
\affil[30]{National Centre for Radio Astrophysics, Post Bag 3, Ganeshkhind, Pune, 411007, India}
\affil[31]{Xinjiang Astronomical Observatory, 150 Science 1-Street, Urumqi, Xinjiang 830011, China}

% Unique acks:
\newcommand{\allacks}{
We thank W. van Straten for help with {\tt digifil}.
The European VLBI Network is a joint facility of independent European, African, Asian, and North American radio astronomy institutes. Scientific results from data presented in this publication are derived from the following EVN project code: EK048.
A.B.P is a McGill Space Institute (MSI) Fellow and a Fonds de Recherche du Quebec - Nature et Technologies (FRQNT) postdoctoral fellow.
B.M. acknowledges support from the Spanish Ministerio de Econom\'ia y Competitividad (MINECO) under grant AYA2016-76012-C3-1-P and from the Spanish Ministerio de Ciencia e Innovaci\'on under grants PID2019-105510GB-C31 and CEX2019-000918-M of ICCUB (Unidad de Excelencia ``Mar\'ia de Maeztu'' 2020-2023).
C.L. was supported by the U.S. Department of Defense (DoD) through the National Defense Science \& Engineering Graduate Fellowship (NDSEG) Program.
D.M. is a Banting Fellow.
E.P. acknowledges funding from an NWO Veni Fellowship.
F.K. acknowledges support from the Swedish Research Council.
FRB research at UBC is supported by an NSERC Discovery Grant and by the Canadian Institute for Advanced Research.
J.P.Y. is supported by the National Program on Key Research and Development Project (2017YFA0402602).
K.S. is supported by the NSF Graduate Research Fellowship Program.
K.W.M. is supported by an NSF Grant (2008031).
M.B. is supported by an FRQNT Doctoral Research Award.
N.W. acknowledges support from the National Natural Science Foundation of China (Grant 12041304 and 11873080).
P.S. is a Dunlap Fellow and an NSERC Postdoctoral Fellow. The Dunlap Institute is funded through an endowment established by the David Dunlap family and the University of Toronto.
V.B. acknowledges support from the Engineering Research Institute Ventspils International Radio Astronomy Centre (VIRAC).
Work at UvA and ASTRON was funded by the NWO Vici grant ``AstroFlash" (PI: Hessels, VI.C.192.045).
}

\date{}
\maketitle
\thispagestyle{empty}
\end{titlepage}

\restoregeometry
\twocolumn
%current main text word count (excluding summary paragraph) is 3493 words

%\section*{Abstract}
\textbf{
%213 words
Fast radio bursts (FRBs) are extragalactic radio flashes of unknown physical origin \citep{petroff_2019_aarv,cordes_2019_araa,petroff_2021_arxiv}.  Their high luminosities and short durations require extreme energy densities, like those found in the vicinity of neutron stars and black holes.  Studying the burst intensities and polarimetric properties on a wide range of timescales, from milliseconds down to nanoseconds, is key to understanding the emission mechanism.  However, high-time-resolution studies of FRBs are limited by their unpredictable activity levels, available instrumentation and temporal broadening in the intervening ionised medium.  Here we show that the repeating \frb\ \citep{bhardwaj_2021_apjl} can produce isolated shots of emission as short as about $60$ nanoseconds in duration, with brightness temperatures as high as $3\times 10^{41}$\,K (excluding relativistic effects), comparable to `nano-shots' from the Crab pulsar. Comparing both the range of timescales and luminosities, we find that \frb\ observationally bridges the gap between known Galactic young pulsars and magnetars, and the much more distant extragalactic FRBs. This suggests a common magnetically powered emission mechanism spanning many orders of magnitude in timescale and luminosity. In this work, we probe a relatively unexplored region of the short-duration transient phase space; we highlight that there likely exists a population of ultra-fast radio transients at nanosecond to microsecond timescales, which current FRB searches are insensitive to.}

%
%-------------------------------------------------------------------
%\section{Introduction}\label{sec:intro}
% 433 words
In the late 1960s, the discovery of radio pulsars was enabled by observations using an unprecedented time resolution of $50-100$\,ms \citep{hewish_1968_natur}.  The subsequent discovery of the Crab pulsar\citep{staelin_1968_sci} eventually led to the insight that its giant pulses \citep{heiles_1970_natur,staelin_1970_natur} show structure on timescales as short as 0.4 nanoseconds \citep{hankins_2003_natur}. These `nano-shots' have exceptionally high brightness temperatures of $\sim10^{40}$\,K \citep{hankins_2007_apj}. The more recent discovery of the fast radio burst (FRB) phenomenon \citep{lorimer_2007_sci,thornton_2013_sci} has shown that nature produces millisecond duration radio flashes that are apparently $10^{9-12}$ times more luminous than those of average pulsars \citep{petroff_2019_aarv,cordes_2019_araa,petroff_2021_arxiv}.  

As with pulsars, the recording of raw voltage data can allow us to study FRB signals on timescales of microseconds (\us) down to nanoseconds (ns).  Thus far, such observations have demonstrated that FRB signals can show temporal structure on timescales of tens of microseconds down to just a few microseconds \citep{farah_2018_mnras,michilli_2018_natur,cho_2020_apjl,nimmo_2021_natas}.  \citet{nimmo_2021_natas} also discuss how individual FRBs can display a range of temporal structure, from a few microseconds up to milliseconds.  The temporal behaviour of FRBs --- along with their polarimetric properties \citep{michilli_2018_natur,day_2020_mnras} and dynamic spectra \citep{hessels_2019_apjl} --- provide key inputs for distinguishing between the dozens of proposed source models and emission mechanisms \citep{platts_2019_phr}.  Furthermore, these burst properties can aid in observationally distinguishing the populations of repeating and apparently one-off FRBs, thereby strengthening the case that they have physically distinct origins \citep{chime_2019_apjl,fonseca_2020_apjl,pleunis_2021_arxiv}.

Kirsten et al.\,(submitted) recently associated the repeating \frb\ \citep{bhardwaj_2021_apjl} with \gc, a globular cluster that is part of the M81 galactic system.  At a distance of 3.6\,Mpc, \frb\ is the closest-known extragalactic FRB, bridging the gap between the putative Galactic FRB source \sgr\ (a known magnetar at a distance of $\sim3-10$\,kpc)\citep{chime_2020_natur_587,bochenek_2020_natur,zhong_2020_apjl} and \rthree\ at a luminosity distance of 149\,Mpc \citep{marcote_2020_natur}. \frb is also at high Galactic latitude ($\sim 41.2^{\circ}$), which reduces the effect of temporal scatter broadening arising from the Milky Way interstellar medium (ISM). This suggests that \frb\ could be an excellent source to study at the highest-possible temporal resolutions.

Here we present a spectro-polarimetric study of 5 \frb\ bursts detected with the Effelsberg 100-m telescope during a broader localisation campaign (Kirsten et al.\,submitted) using \emph{ad hoc} interferometric observations with dishes from the European Very-long-baseline interferometry (VLBI) Network (EVN). In this Article we present both extremely high time resolution, and lower time resolution analyses of the bursts, with full polarimetry, and thereafter discuss the observational connections between \frb\ and well-studied repeating FRBs, the Crab pulsar and the Galactic magnetar \sgr.

\section*{Observations \& Data} \label{sec:data}
% 116 words
We observed \frb\ \citep{bhardwaj_2021_apjl} using an {\it ad hoc} EVN array at 1.4\,GHz, on 2021 February 20 UT 1700 -- 2200, 2021 March 7 UT 1545 -- 2045, and 2021 April 28 UT 1100 -- 2200. For details on the interferometric array configuration and localisation results, see Kirsten et al.\,(submitted). With the 100-m Effelsberg telescope, we recorded dual circular polarization raw voltages (R and L) using 32\,MS/s real sampling per 16-MHz subband and 2-bit samples written in VDIF format \citep{whitney_2010_ivs}, with a total bandwidth of 256\,MHz. 

Throughout this work we label the \frb\ bursts as B$n$, ordered according to their arrival time, and matching the nomenclature used in Kirsten et al.\,(submitted).

\section*{Results}
%1278 words
As described in Methods, we create total-intensity filterbank data containing each burst at the native sampling of the voltage data (31.25\,ns), using SFXC. The filterbank data were created with 32-bit digitization, in order to avoid saturation. The data are coherently dedispersed within the 16\,MHz subbands, and each subband is also time shifted to correct for dispersive delay (incoherent dedispersion), both using our measured dispersion measure (DM) of 87.7527$\pm$0.0003\,\dmunit\ (Methods). We assume the same DM for all bursts. This DM is $> 9 \sigma$ lower than the previously reported measurement of 87.782\,$\pm$\,0.003\,\dmunit\ (where the quoted value is based on the average of 3 bursts) \citep{bhardwaj_2021_apjl}. The observed difference in DM could be due to unresolved time-frequency structure in the \citet{bhardwaj_2021_apjl} bursts, often seen in repeating FRBs \citep{hessels_2019_apjl}. Future measurements are needed to determine if the DM is frequency or time dependent.

In Figure\,\ref{fig:31.25ns}a, we present the profile of burst B3 at 31.25\,ns resolution (black), and downsampled to 1\,$\upmu$s resolution (green), in the frequency range 1318 -- 1334\,MHz. This range corresponds to the single subband containing the brightest spectral feature in the burst (visible in the burst dynamic spectrum shown in Figure\,\ref{fig:bursts}l). By using a single subband, we avoid artefacts due to the inaccuracy of incoherent dedispersion. There are clear few-bin-wide temporal structures in the 31.25\,ns profile of burst B3 (Figure\,\ref{fig:31.25ns}). The question remains whether the sub-microsecond structures we observe are isolated shots of emission or noise fluctuations consistent with the $\chi^2$-distribution of amplitude-modulated noise (AMN; Methods). We calculate the probability of drawing the bright $\sim$\,60\,ns duration feature at Time $=0$ in Figure\,\ref{fig:31.25ns} from the {\it local} brightness distribution, where in this case ``local" is defined as $\pm 1.5625\,\upmu$s ($\pm50$\,bins) around the bright feature (Methods). We find that the probability of drawing this high signal-to-noise (S/N), two-bin-wide structure from the distribution is p\,$=4\times 10^{-8} \times 100\,{\rm bins}/2 = 2\times 10^{-6} $ (Extended Data Figure\,\ref{fig:isol}). Therefore, we find this feature to be inconsistent with the local AMN distribution, and conclude that this structure is a real isolated shot. In addition to this bright 60\,ns shot, we find at least another $2$ significant sub-microsecond shots of emission in burst B3.

Bursts B2 and B4 also have sufficient S/N to study at the highest-possible temporal resolution (Methods; Extended Data Figure\,\ref{fig:isol}). Contrary to burst B3, which exhibits few-bin temporal structure, the highest S/N spikes in B2 and B4 are single-bin unresolved features. This poses a concern since the bright spectral feature dominating the subband used to create the burst profile, has a spectral extent less than the subband width (we attribute the spectral features to scintillation, see below). This results in an effective time resolution less than the native sampling of the data. Therefore, single-bin features are more likely to be consistent with a noise process. The probability of drawing the bright single-bin features from their local distributions are p $=1\times10^{-4} \times 100\,{\rm bins} = 0.01$ and p\,$=4 \times10^{-4} \times 100\,{\rm bins} = 0.04$ for B2 and B4, respectively. These relatively high probabilities, combined with the effective resolution argument above, suggests that the high-resolution features in both B2 and B4 are consistent with the $\chi^2$ AMN distribution.

The total burst duration and spectral structure were quantified by performing a 2-dimensional autocorrelation of the lower-time-resolution burst dynamic spectra (8\,\us, 125\,kHz; Methods). The burst temporal width and frequency extent are reported in Table\,\ref{tab:burst_properties}. As is clear in the autocorrelation functions (ACFs; Extended Data Figure\,\ref{fig:2dacf}), there is an additional narrower frequency scale, which we measure to be consistent with the expected scintillation from the Milky Way ISM \citep{cordes_2002_arxiv} (Methods; Extended Data Figure\,\ref{fig:scintbw}a). The scintillation bandwidth measurements are reported in Table\,\ref{tab:burst_properties}. Additionally, we report the fluence, peak flux density and isotropic-equivalent spectral luminosity of the bursts, computed within the $\pm2\sigma$ width region (Table\,\ref{tab:burst_properties}; Methods).

There are three clear timescales measured in the average temporal ACF (averaged over the four brightest subbands; Methods) of burst B3 at 31.25\,ns resolution (Extended Data Figure\,\ref{fig:acf_b3}f--h): a 40\,\us\ timescale, consistent with the total burst extent in time, a clear 1\,\us\ timescale, and even a shorter timescale (40\,ns) consistent with temporal structure on the few-bin level. In contrast, bursts B2 and B4 both exhibit a timescale on the order of 10\,\us, consistent with their total burst duration (Extended Data Figures \ref{fig:acf_b2} \& \ref{fig:acf_b4}). There is evidence for structure on the few-bin level in the B4 ACF, although the height of this narrow ACF feature relative to the wider ACF feature is smaller for B4 than B3 (height of cyan Lorentzian relative to green Lorentzian in Extended Data Figures \ref{fig:acf_b3}h \& \ref{fig:acf_b4}g), implying either that the S/N of these temporal fluctuations are lower, or that there are fewer temporal features on this timescale. No additional short-timescale components were measured in burst B2 (Extended Data Figure\,\ref{fig:acf_b2}). This further supports the conclusions above that we have resolved microsecond and sub-microsecond structure in burst B3, with no evidence for similar structure in B2 and B4. Additionally, there are a range of timescales observed in the bursts, sometimes observed {\it within} bursts: from tens of nanoseconds to tens of microseconds. In Figure\,\ref{fig:TPS} we compare the isotropic-equivalent luminosity of the three highest-significance shots of emission in the B3 31.25\,ns profile with other short-duration transients, including that of the bright $\sim\,5\,\upmu$s component of B3 seen in the 1\,\us\ resolution data (Extended Data Figure\,\ref{fig:corrcoef}b), and the wider burst envelopes seen in B1, B2, B4 and B5 (Methods; Table\,\ref{tab:burst_properties}).

The power spectra of bursts B2, B3 and B4 (panel \textbf{c}; Extended Data Figures\,\ref{fig:acf_b2}--\ref{fig:acf_b4}) are all consistent with red noise (Methods), with B2 and B4 exhibiting a steeper power law ($\alpha = 1.85\pm0.04$ and $2.04\pm0.05$, respectively) than B3 ($\alpha = 1.46\pm0.05$). This is consistent with the results we have presented above; the power spectrum of B3 shows more power at higher frequencies (shorter timescales), than bursts B2 and B4. 

As described in Methods, the polarimetric data were calibrated using the known polarization properties of the pulsar PSR~B0355+54 \citep{taylor_1993_apjs}. The full polarimetric burst profiles and polarisation position angles (PPA) are shown in Figure\,\ref{fig:bursts}. Note that we could not recover the polarimetric properties of burst B5, likely due to the low S/N of the burst. Bursts B1 -- B4 are highly linearly polarized ($> 90$\%), and exhibit little-to-no circular polarization (Table\,\ref{tab:pol_properties}). As reported in Table\,\ref{tab:pol_properties}, there is a tentative 3 -- 4\,$\sigma$ detection of 13\,\% and 6\,\% circular polarization in B2 and B4, respectively. In the 8\,\us\ resolution profiles, there is evidence for small variations in the PPA across the bursts, with a $\Delta$PPA between bursts from the same epoch within $\sim 30^\circ$. We determine the rotation measure (RM) of the bursts, and conclude that they are in agreement with previous measurements ($-29.8$\,rad\,m$^{-2}$)\citep{bhardwaj_2021_apjl}. 

As presented above, burst B3 exhibits variations on both microsecond as well as sub-microsecond timescales. The Stokes parameters are only physically meaningful with sufficient averaging \citep{vanstraten_2009_apj}. Therefore, in Extended Data Figure\,\ref{fig:B3_125ns_fullpol} we show the frequency-averaged polarization profile and PPA of burst B3 at 125\,ns resolution. In these data we average over a total of 44 channels (over the frequency range 1254 -- 1430\,MHz), and therefore have 44 degrees of freedom. The $\sim$100\,ns -- 1\,$\upmu$s structures are highly linearly polarized, consistent with the polarization properties at lower time resolution. Additionally, the PPA varies between the sub-\us\ temporal features by up to a few 10s of degrees.

\section*{Discussion} 
% 60 words
The timescales and luminosities measured in the \frb\ bursts presented in this work populate a previously vacant, relatively unexplored region of the short-duration transient phase space.  They bridge the gap between extragalactic FRBs and Galactic neutron stars (Figure\,\ref{fig:TPS}). Here we elaborate on how \frb\ compares observationally with other short-duration radio transients, and discuss the implications of our findings. 

\subsection*{Repeating fast radio bursts}
% 499 words
The polarimetric properties of \frb\ are consistent with those of most well-studied repeaters \citep{michilli_2018_natur,nimmo_2021_natas,day_2020_mnras}.  As is often observed, we find a very high ($\sim100$\%) linear and low ($\lesssim\,10$\%) circular polarization fraction as well as a polarization position angle (PPA) that is roughly constant during bursts.  As with \rthree\ \citep{nimmo_2021_natas}, we find that \frb\ shows small PPA  variations ($\Delta$PPA) on timescales $<10$\,\us. Between bursts, we see variations in the PPA of a few tens of degrees, in contrast to the $<10^{\circ}$ $\Delta$PPA from bursts detected at the same observing epoch, seen in other repeaters \citep{michilli_2018_natur,nimmo_2021_natas}. We note that at least one repeater (FRB~20180301A) has shown lower linear polarization fractions ($40-80$\%) and significant PPA swings in some bursts \citep{luo_2020_natur}. 

The spectrum of \frb\ shows at least two scales of variation. We find narrow-band brightness variations on the scale of $\sim 5$\,MHz, which we ascribe to scintillation in the Milky Way ISM. The other $\sim100$\,MHz spectral variation may be intrinsic to the source emission mechanism or imparted by local propagation effects.  This spectral envelope is similar to what is seen from other repeaters \citep{hessels_2019_apjl,gourdji_2019_apjl,pleunis_2021_arxiv}. Additionally, the downward-drifting burst sub-structure, often referred to as the `sad-trombone effect', that is often seen in repeaters \citep{hessels_2019_apjl}, has previously been observed for \frb\ \citep{bhardwaj_2021_apjl}. 

Though \frb\ shares many characteristics of repeating FRBs, its $\sim\,100$\,\us\ burst envelopes are atypically narrow. The range of timescales observed, however, roughly a factor of 1000, is comparable to what has been found in a similar analysis of \rthree\ \citep{nimmo_2021_natas}. In the case of \rthree, \citet{nimmo_2021_natas} were limited by a larger scatter-broadening of 2.7\,\us\ (note that \rthree\ is at a Galactic latitude of only $3.7^{\circ}$; see also Extended Data Figure~\ref{fig:scintbw}b) \citep{marcote_2020_natur}, and could not rule out the possibility that the wider sub-bursts are composed of closely spaced microsecond structures. Burst B3 from \frb\ has clear isolated shots of duration $\sim\,60$\,ns, and also evidence that the sub-microsecond shots are clustered on microsecond timescales (Extended Data Figure\,\ref{fig:acf_b3}f--h). For bursts B2 and B4, there is no clear evidence for isolated shots in the 31.25\,ns burst profile, but it is possible that the S/N is too low to detect these individual shots, consistent with the lower $<$0.33 measured correlation coefficient (Methods; Extended Data Figure\,\ref{fig:corrcoef}d--f).

The strikingly similar observational properties of \frb\ with other repeaters suggests that they have similar physical origins. \frb\ is the closest known extragalactic FRB discovered to date (Kirsten et al. submitted). The remarkable proximity of \frb\ has revealed radio bursts with an isotropic-equivalent spectral luminosity 2--3 orders of magnitude weaker than bursts from other repeating FRB (Figure\,\ref{fig:TPS}). Such low luminosity bursts would not be detectable at the distance of any other precisely localized repeating FRBs. Continued monitoring of \frb\ will be important to compare the energy distribution and activity rate with other repeaters.

\subsection*{Crab pulsar}
% 387 words
We have discovered resolved structure in bursts from \frb\ with durations of $\sim\,5$\,\us, down to $\sim$\,60\,ns, $2$ orders of magnitude shorter timescales than have been probed for FRBs, to date. Crab pulsar giant pulses (GPs) show temporal structure in the range \us\ -- ms, which sometimes resolve down to sub-nanosecond timescales, often referred to as `nano-shots' \citep{hankins_2003_natur}. Crab nano-shots are known to be extremely energetic, with brightness temperatures reported up to $10^{41}$\,K (ignoring relativistic effects) \citep{hankins_2007_apj}. The sub-microsecond shots in B3 exhibit a comparable, extremely high brightness temperature (again, ignoring relativistic effects; Figure~\ref{fig:TPS}). This is at least four orders of magnitude higher than typically seen from the FRB population ($10^{32}$--$10^{37}$\,K) \citep{petroff_2019_aarv}. In the line of sight to \frb, the expected scatter broadening from the Milky Way ISM is $\sim50$\,ns at 1.4\,GHz \citep{cordes_2002_arxiv}, in rough agreement with our measured scintillation bandwidth ($1/(2\pi\Delta\nu_{\rm scint})\sim 27$\,ns). This implies that the $\sim\,60$\,ns structure observed in this work is likely the shortest resolvable temporal scale at our observing frequency. 

Bursts B2, B3 and B4 are consistent with the Scintillating Amplitude Modulated Polarized Shot Noise model (Methods), which has been used to describe many aspects of pulsar emission, including GPs from the Crab pulsar \citep{cordes_2004_apj,karuppusamy_2010_aa}. However, only B3 is observed to resolve down to microsecond and sub-microsecond temporal scales. This is similar to observations of the Crab pulsar, where not all GPs resolve down to individual nano-shots \citep{hankins_2007_apj,jessner_2010_aa}. Furthermore, most Crab GPs consist of at least one broadband `micro-burst', with a characteristic timescale of a few microseconds \citep{hankins_2007_apj,hankins_2016_apj}, similar to the structure that can be seen in burst B3 at 1\,\us\ time resolution and consistent with the measured 1.11\,\us\ timescale in the temporal ACF (Extended Data Figures\,\ref{fig:acf_b3}g \& \ref{fig:corrcoef}b). Sometimes, the Crab micro-bursts are seen to resolve down to individual narrowband nano-shots (see e.g. Figure 4 of \citet{hankins_2007_apj}), consistent with the features seen at 31.25\,ns in burst B3.

Band-limited giant pulses from the Crab pulsar\citep{hankins_2007_apj,thulasiram_2021_arxiv, bij_2021_arxiv}, as well as from the `Crab's twin' pulsar PSR~J0540$-$6919 \citep{geyer_2021_mnras}, have been observed.  These are reminiscent of the narrow-banded emission of observed from repeating FRBs. This adds further weight to the phenomenological connection of giant pulse emission with FRBs. 

\subsection*{Galactic magnetar SGR~1935$+$2154}
% 140 words
The \frb\ bursts also exhibit wider characteristic timescales of $10-100$\,\us. These wider components have isotropic-equivalent energies on the order of $10^{32-33}$\,erg, which is 2--3 orders of magnitude weaker than the bright FRB-like radio burst from the Galactic magnetar SGR~1935+2154 \citep{bochenek_2020_natur,chime_2020_natur_587}. Thus, SGR~1935+2154 has produced more energetic radio bursts than some extragalactic FRBs, eliminating the gap in luminosity (Figure\,\ref{fig:TPS}) and thereby strengthening the connection between magnetars and FRBs. The observed energies are much lower than the proposed low-energy cutoff at $\sim$10$^{34}$\,erg for the low-twist magnetar model \citep{wadiasingh_2020_apj}, implying that the FRB luminosity function does not abruptly end at $\sim$10$^{34}$\,erg. The brightness temperatures of these wider components are $\sim 10^{32}$\,K, comparable to \sgr\  \citep{bochenek_2020_natur,chime_2020_natur_587}, and also consistent with the lower end of the observed FRB brightness temperatures (Figure~\ref{fig:TPS}).

\subsection*{Implications}
% 580 words
Previously, it has been shown that millisecond-duration radio bursts from the Galactic magnetar SGR~1935+2154 span 7--8 orders of magnitude in apparent luminosity \citep{kirsten_2021_natas}, bridging from the brightest pulsars, to 1--2 orders of magnitude lower than the weakest known extragalactic FRBs. By probing \frb\ at timescales of tens of nanoseconds, we highlight that this coherent radio source observationally ties repeating FRBs with GP emission from young pulsars, and the bright FRB-like emission from magnetars. The timescales and luminosities measured in this work fill the gap in the luminosity-duration phase space (Figure\,\ref{fig:TPS}), further emphasising that observationally the division between source populations (FRBs, pulsars, magnetars) is not clear. 

Constraints on the shortest timescale variations in FRB lightcurves are key for understanding the physical mechanism producing the bursts, and can ultimately reveal clues about the progenitor. The observational connection to the Crab pulsar, \sgr, and repeating FRBs, supports a common magnetically-powered emission mechanism spanning many orders of magnitude in timescale and luminosity. The $\sim$\,60\,ns to 5\,\us\ timescales observed in B3 correspond to a light-travel size of 20--1500\,m, ignoring relativistic effects.  The sub-microsecond timescales observed are too short to be naturally explainable via emission from a synchrotron maser in a relativistic shock \citep{metzger_2019_mnras}, since it would require a small region to be emitting at any given time.  \citet{nimmo_2021_natas} previously argued that the short timescales observed in \rthree\ are more naturally explainable in the context of a neutron star magnetospheric origin. The results presented here further support a magnetospheric origin of the FRB emission. 

Magnetic reconnection is one possible way to harness the magnetic energy to power a coherent radiation mechanism that produces the radio emission. This has been proposed to explain Crab nano-shots \citep{philippov_2019_apjl}, the FRB-like radio burst from \sgr\ \citep{yuan_2020_apjl}, and FRBs \citep{lyubarsky_2020_apj,lyutikov_2021_arx}. A range of temporal scales and magnetic energy releases can be expected from this mechanism, which is consistent with the observed dynamic range of timescales, and large range of FRB luminosities; from the weaker possible FRBs emitted by the relatively old magnetar SGR~1935+2154, and the bursts from \frb, to the more energetic FRBs potentially coming from extragalactic young, active magnetars.

The observed timescales and luminosities from \frb\ can be explained by magnetic reconnection events in the close vicinity of a relatively young, isolated, highly magnetized neutron star. The association of \frb\ with an old globular cluster strongly implies that, if the source is a magnetar, it was not formed through a core-collapse supernova (Kirsten et al. submitted). The globular cluster origin of \frb\ also allows for the exploration of alternatives to a magnetar progenitor: for example, an accreting system, in which case the observational similarities with the Crab pulsar and \sgr\ are more coincidental. 

Future observations of \frb\ at observing frequency $\geq 5$\,GHz (where the scatter broadening will be lower), with a bandwidth $>200$\,MHz, are needed to explore shorter timescales. Continued monitoring of \frb\ will provide statistics on the distribution of emission timescales and whether the burst activity is periodic, like \rthree\ \citep{chime_2020_natur_582}.  Such data will also help determine whether, e.g., there are quasi-periodic fluctuations in the burst lightcurves, hinted at in the case of \rthree\ \citep{nimmo_2021_natas}, or a secular variation in the widths of burst envelopes. 

Lastly, we highlight that the short timescales measured from \frb\ in this work motivate searches for a population of {\it ultra}-fast radio bursts, despite the considerable technical challenges that such a search entails. 

\begin{figure*}
\resizebox{\hsize}{!}
        {\includegraphics[trim=0cm 1cm 0cm 2cm, clip=true]{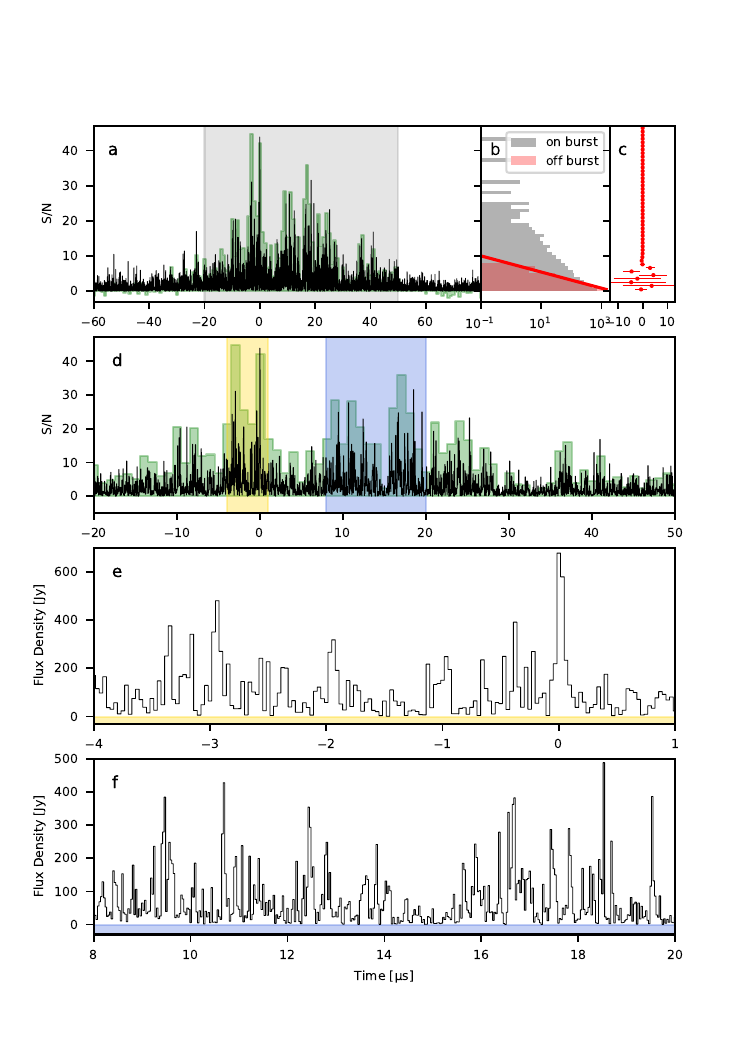}}
  \caption{Burst B3 from \frb\ exhibits sub-microsecond temporal structure. Panel \textbf{a} shows the temporal profile of burst B3 at 31.25\,ns time resolution (black) and downsampled to 1\,\us\ resolution (green). Panel \textbf{b} shows a histogram of the S/N values in the 31.25\,ns data (black), also showing the $\chi^2$-distributed (2 degrees of freedom) noise distribution (red) arising in the limiting case of $\delta t \delta \nu \sim 1$, where $\delta t$ and $\delta \nu$ are the time and frequency resolutions of the data, respectively. The residuals of the off-burst histogram with the best-fit $\chi^2$-distributed with 2 degrees of freedom is shown in panel \textbf{c}. This plot contains temporal profiles computed for the bright scintle only, i.e. the frequency range $1318-1334$\,MHz. The data were generated with {\tt SFXC} and are coherently dedispersed to 87.7527\,\dmunit. Panel \textbf{d} shows a zoom in of panel \textbf{a}, highlighting the temporal regions plotted in panels \textbf{e} (yellow) and \textbf{f} (blue) panels.}
     \label{fig:31.25ns}
\end{figure*}

\begin{figure*}
\resizebox{\hsize}{!}
        {\includegraphics[trim=0cm 0cm 0cm 0cm, clip=true,width=\textwidth,height=195mm]{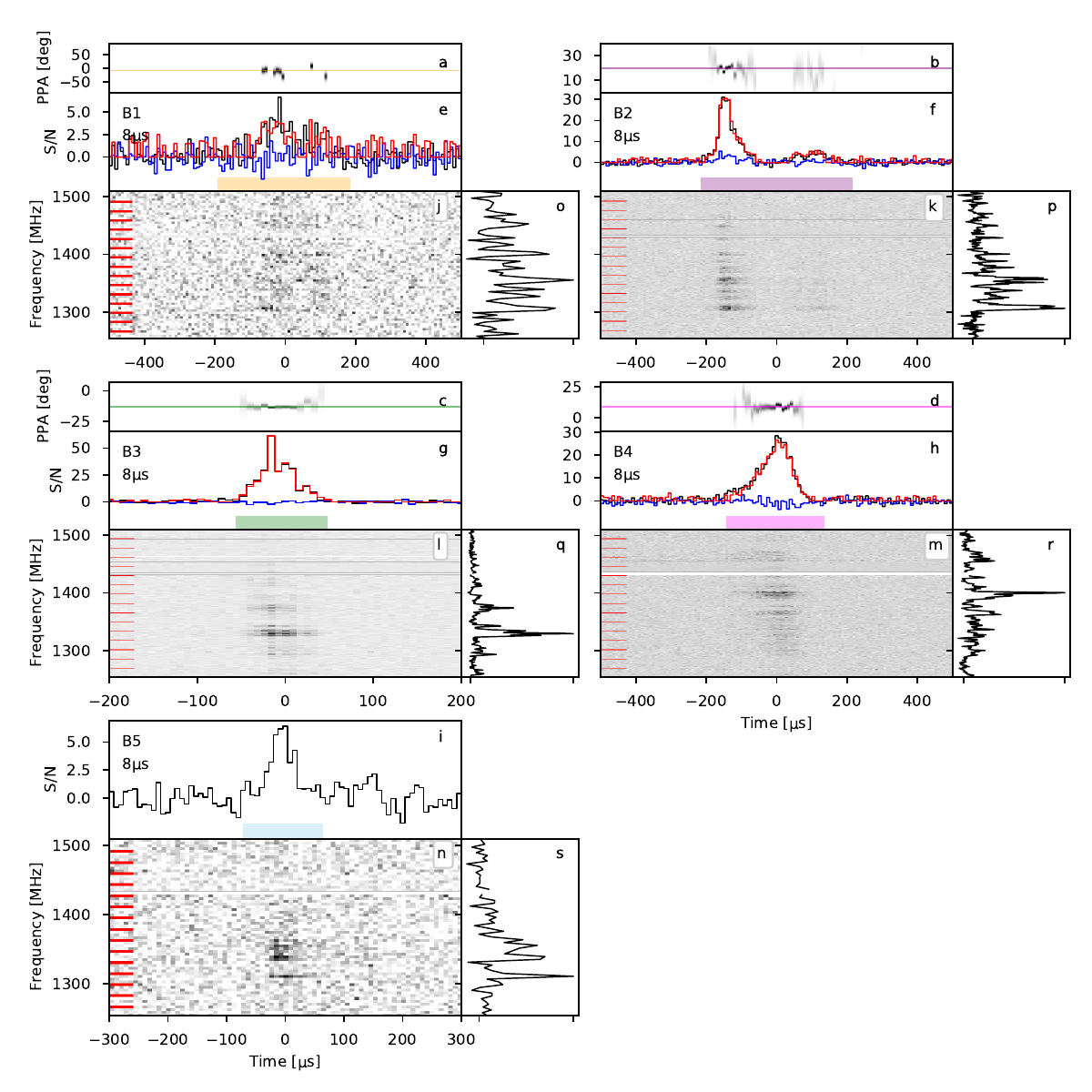}}
  \caption{The polarimetric profiles, dynamic spectra, time-averaged spectra and polarization position angle (PPA) of the bursts detected from \frb. For B5 only Stokes I is shown (see text). The data are plotted with 8\,$\upmu$s and 1\,MHz time and frequency resolution, respectively (with the exception of B1 and B5 which are plotted with 4\,MHz frequency resolution). The data were generated with {\tt SFXC} and are coherently dedispersed within each 16\,MHz subband to 87.75\,\dmunit (and also incoherently shifted between subbands). Panels \textbf{a}--\textbf{d} are the PPA across the burst profile, where the color gradient represents the linear polarization S/N (black is high S/N and white is low S/N), and the colored horizontal line represents the weighted best-fit line to the PPA. Only the PPAs above a linear S/N threshold of 3 are plotted. Panels \textbf{e}--\textbf{h} show the total intensity (Stokes I; black), unbiased linear polarization (\citealt{everett_2001_apj}; red) and circular polarization (blue) burst profiles (panel \textbf{i} shows the Stokes I profile of burst B5). In the top-left of the panels are the burst name used throughout this work, and the time resolution used for plotting. The colored bar at the bottom of the panels represent the $\pm$2$\sigma$ burst width used to measure the polarization fractions and burst fluence. Panels \textbf{j}--\textbf{n} are the dynamic spectra and panels \textbf{o}--\textbf{s} are the time-averaged frequency spectra. The red marks on the dynamic spectra outline the edges of the subbands. Data that have been removed due to radio frequency interference have not been plotted.}
     \label{fig:bursts}
\end{figure*}

\begin{figure*}
\resizebox{\hsize}{!}
        {\includegraphics[trim=0cm 0cm 0cm 1cm, clip=true]{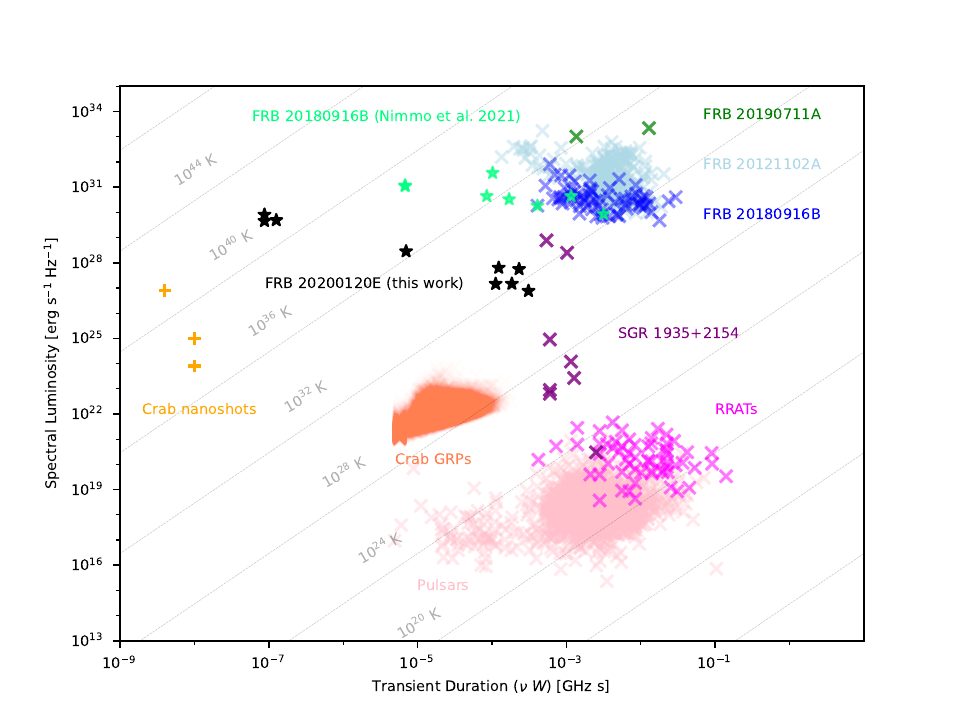}}
  \caption{Nanosecond to second transient phase space. The \frb\ burst temporal structures and their measured isotropic-equivalent spectral luminosity are shown by the black stars. We plot the three highest significance isolated shots from burst B3 (left-most black stars), the bright $5$\,\us\ structure evident in the 1\,\us\ time resolution data of burst B3 (Extended Data Figure\,\ref{fig:corrcoef}), as well as the wider burst structure seen in B1, B2, B4 and B5 (right-most black stars). The other FRBs plotted are the published localized repeating FRBs, with known distances: FRB~20121102A \citep{spitler_2016_natur,scholz_2017_apj,law_2017_apj,michilli_2018_natur,hessels_2019_apjl,gourdji_2019_apjl,gajjar_2018_apj,hardy_2017_mnras,houben_2019_aa,majid_2020_apjl,josephy_2019_apjl,rajwade_2020_mnras,caleb_2020_mnras}, FRB~20180916B \citep{chime_2019_apjl,marcote_2020_natur,chawla_2020_apjl,chime_2020_natur_582,pleunis_2021_apjl,nimmo_2021_natas}, and FRB~20190711A \citep{macquart_2020_natur,kumar_2021_mnras}. In purple we plot the radio bursts from the Galactic magnetar SGR~1935+2154 \citep{chime_2020_natur_587,bochenek_2020_natur,zhang_2020_atel,kirsten_2021_natas,good_2020_atel}. Giant pulses from the Crab pulsar (GRPs) are shown in orange \citep{karuppusamy_2010_aa}, and the `nano-shots' are represented by the yellow crosses \citep{hankins_2003_natur,hankins_2007_apj,jessner_2010_aa}. The pulsar and Rotating RAdio Transient (RRAT) population are shown in pink \citep{keane_2018_natas}. Note that for pulsars, the points on the figure represent average pulses, where individual pulses can be brighter or weaker by approximately 1 or 2 orders of magnitude. The grey lines represent lines of constant brightness temperature.}
     \label{fig:TPS}
\end{figure*}

\begin{table*}
\caption{\label{tab:burst_properties}Burst properties.}
\resizebox{\textwidth}{!}{\begin{tabular}{lcccccccc}
\hline
\hline
		{Burst}& {Time of Arrival$\mathrm{^{a}}$ } & {Fluence$\mathrm{^{b}}$} & {Peak S/N$\mathrm{^{c}}$} & {Peak Flux Density$\mathrm{^{b,c}}$} & {Spectral Luminosity$\mathrm{^{b,d,e}}$} & {Width $\mathrm{^{f}}$ } & {Frequency Extent $\mathrm{^{f}}$ } & {Scintillation bandwidth$\mathrm{^{h}}$ } \\
		{} & {[MJD]} & {[Jy m0s]} & &{[Jy]} & {[10$^{28}$\,erg\,s$^{-1}$\,Hz$^{-1}$]} & {[$\upmu$s]} & {[MHz]} & {[MHz]}\\
		\hline
		B1 &  59265.88304437179   & 0.13\,$\pm$\,0.03 & 6.6 & 1.59\,$\pm$\,0.32 & 0.56\,$\pm$\,0.13 & 156 $\pm$ 1 & 140\,$\pm$1 & 1.9 $\pm$ 0.7  \\
        B2 &  59265.88600912486  & 0.63\,$\pm$\,0.13 & 36.1 & 8.71\,$\pm$\,1.74 & 2.3\,$\pm$\,0.6 & 62\,$\pm$\,1, 93 $\pm$ 0.5 $\mathrm{^{g}}$ & 103\,$\pm$\,1, 89\,$\pm$\,1 $\mathrm{^{g}}$ & 3.4 $\pm$ 1.0 \\
        B3 & 59280.69618745651  & 0.53\,$\pm$\,0.11 & 64.8 & 15.6\,$\pm$\,3.12 & 8.0\,$\pm$\,1.9 & 46.7 $\pm$ 0.1 & 94\,$\pm$\,1 & 6.9 $\pm$ 1.1 \\
        B4 & 59280.80173397988  & 0.71\,$\pm$\,0.14  & 29.3 & 7.07\,$\pm$\,1.41 & 4.0\,$\pm$\,1.0 & 117\,$\pm$\,1 & 134\,$\pm$1 & 6.8 $\pm$ 1.3 \\
        B5 & 59332.50446581106 & 0.09\,$\pm$\,0.02 & 6.9 & 1.66\,$\pm$\,0.33 & 1.0\,$\pm$\,0.2 & 56.6\,$\pm$\,0.1 & 86\,$\pm$\,1 & 2.9\,$\pm$\,1.5\\
		\hline
		\multicolumn{9}{l}{$\mathrm{^{a}}$ Corrected to the Solar System Barycenter to infinite frequency assuming a dispersion measure of $87.75$ pc\,cm$^{-3}$,}\\
		\multicolumn{9}{l}{ \hspace{0.3cm} reference frequency 1502\,MHz and dispersion constant of 1/(2.41$\times 10^{-4}$)\,MHz$^2$\,pc$^{-1}$\,cm$^{3}$\,s.} \\
		\multicolumn{9}{l}{ \hspace{0.3cm} The times quoted are dynamical times (TDB).}\\
		\multicolumn{9}{l}{$\mathrm{^{b}}$ We estimate a conservative 20\% error on these measurements, arising due to the uncertainty in the system equivalent flux density (SEFD) of Effelsberg.} \\
		\multicolumn{9}{l}{$\mathrm{^{c}}$ Determined for a time resolution of 8\,\us.}\\
		\multicolumn{9}{l}{$\mathrm{^{d}}$ Using the distance to the globular cluster \gc\ of 3.63\,$\pm$0.34\,Mpc \citep{freedman_1994_apj}.} \\
		\multicolumn{9}{l}{$\mathrm{^{e}}$ Isotropic-equivalent.} \\
        \multicolumn{9}{l}{$\mathrm{^{f}}$ Defined as $1/\sqrt{2}$ multiplied by the FWHM} \\
		\multicolumn{9}{l}{ \hspace{0.3cm} of the autocorrelation function (ACF).}\\
		\multicolumn{9}{l}{$\mathrm{^{g}}$ Width per burst component.} \\
		\multicolumn{9}{l}{$\mathrm{^{h}}$ The uncertainties are a combination of the 1-$\sigma$ fit uncertainty, and 1/$\sqrt{\rm N}$, where N\,$\approx$\,5 is the number of scintles.}\\
		\end{tabular}}
        
\end{table*}

\begin{table*}
\caption{\label{tab:pol_properties}Burst polarization properties and polarization position angle fit results.}
\resizebox{\textwidth}{!}{\begin{tabular}{lccccccccc}
\hline
\hline
		{Burst} & {RM$^{\rm a}$ [rad m$^{-2}$]} & {PPA offset$^{\rm b}$ [deg]} & {RM (global)$^{\rm a,c}$ [rad m$^{-2}$]} & {PPA offset (global)$^{\rm c}$ [deg]} & {L$_{\rm unbiased}$/I$^{\rm d}$ [\%]} & {V/I$^{\rm d}$ [\%]}  & $\chi^{2}$ $^{\rm e}$ & Degrees of freedom$^{\rm e}$ \\
		\hline
		B1 & $-$21.9 $\pm$ 13.1 & 28.5 & $-$54.2 $\pm$ 4.8 & 68.3 & 94 $\pm$ 9 & 4 $\pm$ 9 &    18.9 & 7 \\
        B2 & $-$57.2 $\pm$ 5.1 & 0 & $-$54.2 $\pm$ 4.8 & 0 & 103 $\pm$ 3 & $-$13 $\pm$ 3 &  59.6 & 27 \\
        B3 & $-$37.1 $\pm$ 4.2 & $-$8.6 & $-$36.9 $\pm$ 3.8 & $-$9.3 & 102 $\pm$ 2 & 1 $\pm$ 1 & 15.5  & 11 \\
        B4 & $-$36.4 $\pm$ 8.0 & 13.1 & $-$36.9 $\pm$ 3.8 & 14.3 & 94 $\pm$ 2 & 6 $\pm$ 2 &   25.7 & 22 \\
        \hline
		\multicolumn{8}{l}{$\mathrm{^{a}}$ The large fractional errors arise due to covariances between fit parameters in the QU-fitting, see text for more details.}\\
		\multicolumn{8}{l}{$\mathrm{^{b}}$ Offset from a weighted mean of the bursts from a single epoch.}\\
		\multicolumn{8}{l}{$\mathrm{^{c}}$ The global values are assuming all bursts detected on the same day have the same RM. }\\
		\multicolumn{8}{l}{$\mathrm{^{d}}$ The quoted uncertainties are 1$\sigma$ uncertainties, and do not include uncertainties from the calibration.}\\
	    	\multicolumn{8}{l}{\hspace{0.3cm} Note: we reproduce the published pulsar circular (linear) fractions within 3\% (10\%) and 1\% (6\%)}\\
		    \multicolumn{8}{l}{\hspace{0.3cm} for the Feb 20 and March 7 observations, respectively. }\\
		\multicolumn{8}{l}{$\mathrm{^{e}}$ Fit results of weighted least squares fitting of a straight line to the PPA.}\\
        
\end{tabular}}
        
\end{table*}

\newpage
\section*{Methods}
In this section, we describe the analysis performed in more detail. Since this was an interferometric campaign, we regularly observed a phase calibrator source (5.5\,minutes on target, 2.0\,minutes on the calibrator). The total time on \frb\ was 2.93\,hr per observation. 
\subsection*{Burst search}
We performed two independent analyses on the single-dish Effelsberg data to search for bursts from \frb. The voltage data were converted to total-intensity filterbanks using {\tt digifil} with time and frequency resolutions of 64\,$\upmu$s and 125\,kHz, respectively. These data were searched using a {\tt Heimdall}-based pipeline and burst candidates were classified using the machine learning classifier, FETCH \citep{agarwal_2020_mnras}. A more detailed description of this pipeline was presented in \citet{kirsten_2021_natas}. In parallel, we also recorded total intensity filterbanks using the PSRIX pulsar backend \citep{lazarus_2016_mnras}, with time and frequency resolution of 102.4\,$\upmu$s and 0.49\,MHz, respectively, and a total bandwidth of 250\,MHz. The PSRIX pulsar backend data were analysed using a PRESTO-based pipeline \citep{ransom_2001_phdt}, and single pulse candidates were classified using an automated classifier based on \citet{michilli_2018_mnras}. We recorded a test pulsar scan of PSR~B0355+54 to inspect the data quality. 

Five bursts from \frb\ were discovered in the search of the raw voltages using the {\tt Heimdall}-based search (B1 and B2 on 2021 February 20; B3 and B4 on 2021 March 7; and B5 on 2021 April 28). Only four of the five bursts were found in the PSRIX data because burst B1 occurred during the $\sim$\,40\,s latency between the start of the VLBI recording and the pulsar backend recording. Initially, B2 was flagged as radio frequency interference in the PSRIX data by the automated classifier \citep{michilli_2018_mnras} due to strong scintillation structure. 
\subsection*{Data products}
For further analysis of the burst properties — using full polarimetric information at a range of time and frequency resolutions — we generated filterbank data from the raw voltages using both SFXC \citep{keimpema_2015_exa} and {\tt digifil} \citep{vanstraten_2011_pasa}. 

\subsubsection*{SFXC}

Using SFXC (phased array branch\footnote{https://github.com/aardk/sfxc/tree/phased-array}), we produced relatively low time resolution filterbank data (8\,$\upmu$s), in order to have sufficient spectral resolution (125\,kHz) to study frequency structure due to scintillation, and to determine the rotation measure (RM). The data are coherently dedispersed within the 16-MHz subbands, and incoherently dedispersed between the subbands using a DM of $87.75$\,pc\,cm$^{-3}$, which is within $\sim 0.003$\,pc\,cm$^{-3}$ of the value we measured using higher time resolution data (see below). This $0.003$\,pc\,cm$^{-3}$ offset corresponds to a dispersive delay, across the burst extent in frequency, that is less than the time resolution. Using {\tt dspsr} \citep{vanstraten_2011_pasa} we created archive files containing each burst at the resolution of the filterbank data, and remove frequency channels from the data that are contaminated by radio frequency interference using the {\tt psrchive} tool {\tt pazi} \citep{hotan_2004_pasa}.

The real-sampled VDIF data with 16-MHz subbands allow for up to 31.25\,ns time resolution. Due to the Fast Fourier Transform-based (FFT-based) correlation and delay corrections implemented in SFXC (phased array), the practical time resolution limit is 125\,ns. In this case, the data have 4 channels per subband. Since SFXC does not window the data before the FFT, potential spectral leakage artefacts are a concern, especially in the case of a low number of channels per subband. This data is used to study the polarimetry at high time resolution.

Additionally, we output coherently dedispersed total intensity (Stokes I) at the original data sampling rate (i.e. no channelization within subbands) using the SFXC bolometer branch\footnote{https://github.com/aardk/sfxc/tree/bolometer}. Here we have no concerns of spectral leakage or any other artefacts that may be introduced at the channelization stage of the processing. Note that the VLBI backed, DBBC2 \citep{tuccari_2010_ivs}, uses digital down conversion to create independent 16\,MHz subbands, meaning that adjacent 31.25\,ns samples are not correlated.

\subsubsection*{Digifil}

We also used {\tt digifil} to produce data at the highest-possible time resolution, in order to verify the results presented in this work. {\tt digifil} utilises a convolving filterbank, which reduces spectral leakage effects \citep{vanstraten_2011_pasa}. However, {\tt digifil} converts to complex-sampled data internally, resulting in a factor of 2 decrease in the time resolution for a given number of channels (for the Nyquist limit). Thus, the best time resolution we can achieve with our 16-MHz subbands is 62.5\,ns. {\tt digifil} produces one filterbank file per subband, and the subbands are combined with incoherent dedispersion. 

In both the {\tt digifil} and SFXC data, the incoherent dedispersion between subbands is performed with integer sample shifts.

\subsection*{Refined dispersion measure determination}
At the native time resolution of the voltage data (31.25\,ns), the DM must be correct to $< 0.0005$\,\dmunit\ to avoid DM smearing across the 256\,MHz bandwidth larger than the time bin width.

To limit potential spectral leakage artefacts in the SFXC data, we produced 500\,ns time resolution filterbank files, with 16 channels per subband (1\,MHz channels), at a range of DM values around the expected DM (determined by eye in the dynamic spectrum of burst B3). The data are coherently dedispersed within the 16\,MHz subbands, and each subband is then time shifted to correct for dispersive delay, to limit DM smearing within each channel, and across the band. These data were then downsampled in time to 1\,$\upmu$s (i.e. a downsampling factor of 2). The reason for producing the data products in this way, as opposed to creating 1\,$\upmu$s resolution data with higher frequency resolution, is to minimise smearing due to incoherent dedispersion between subbands (integer bin shift). We can accurately determine the DM of burst B3 by maximizing the S/N of the bright $\sim\,5$\,\us\ component (Extended Data Figure\,\ref{fig:DM}), despite any downward-drifting, `sad-trombone' effect, that may be present in the burst dynamic spectrum. This is because by maximizing the S/N of the narrow temporal structure, we are essentially maximizing the structure in the burst (as opposed to maximizing the S/N of the entire burst, which has been shown to incorrectly capture the DM, e.g. \citealt{gajjar_2018_apj}). In Extended Data Figure~\ref{fig:DM}, we plot the peak S/N of the burst profile as a function of DM, and fit a Gaussian function to determine the best DM. We search for a DM in the range $87.74-87.765$\,\dmunit\ in steps of 0.001\,\dmunit\ and find that the best DM is DM$=87.7527\,\pm\,0.0003$\,\dmunit, where the uncertainty is determined by $\sigma/\mathrm{A}$, where $\sigma$ and $A$ are the standard deviation and amplitude of the Gaussian fit, respectively. We note that the range of DMs searched do not include the previously measured DM \citep{bhardwaj_2021_apjl}, since the dynamic spectrum of burst B3 is visibly over-corrected when dedispersed to this value. Both measurements of the DM use the same dispersion constant of 1/(2.41$\times 10^{-4}$)\,MHz$^2$\,pc$^{-1}$\,cm$^{3}$\,s, thus they can be directly compared. 

In panels \textbf{c} and \textbf{d} of Extended Data Figure~\ref{fig:DM} we plot the burst profile and dynamic spectrum coherently (within channels) and incoherently (between subbands) dedispersed to this best-fit DM, respectively. This analysis was repeated on {\tt digifil} filterbank data products generated with 62.5\,ns time and 16\,MHz frequency resolution, coherently and incoherently dedispersed, and then downsampled to 1\,\us\ time resolution. We find a consistent DM value from {\tt digifil} and SFXC data (noting also that both softwares use the same dispersion constant of 1/(2.41$\times 10^{-4}$)\,MHz$^2$\,pc$^{-1}$\,cm$^{3}$\,s). 

We assume the same DM for all bursts in our sample.

\subsection*{Scintillating Amplitude Modulated Polarized Shot Noise}

In the uncertainty principle limit $\delta t \delta \nu \sim 1$, where $\delta t$ and $\delta \nu$ are the time and frequency resolution (respectively) of individual samples, the off-burst noise is $\chi^2$-distributed with 2 degrees of freedom (Figure \ref{fig:31.25ns}b). For a modestly broadband noise-like signal, the statistics are $\chi^2$-distributed where the degrees of freedom depend on the polarization fraction. We perform a least-squares fit of a $\chi^2$-distribution to the on-burst distribution and find the best fit to have 2 degrees of freedom, consistent with 100\,\% polarized Scintillating Amplitude Modulated Polarized Shot Noise (SAMPSN) \citep{cordes_1976_apj}. The superposition of the off-burst $\chi^2$-distribution and the on-burst $\chi^2$-distribution reasonably describes the total burst S/N distribution, where the residuals can likely be attributed to the fact that the shape of the burst envelope is not well-modelled. 

Due to strong spectral dips near the subband edges, sufficient frequency resolution is required in order to robustly correct the bandpass and compare individual bin spectra with each other. We therefore created a 1\,\us, 500\,kHz filterbank using SFXC. Shown in Extended Data Figure~\ref{fig:corrcoef} is the correlation coefficient between individual 1\,\us\ time bins above a S/N threshold of 9 in burst B3, as a function of their time separation. We find that the correlation coefficients have a geometric mean S/N weighted average of 0.31.
The scintillation frequency structure is expected to be perfectly correlated within the duration of the burst (we observe that the B1 and B2 spectra, separated by 4.3\,minutes, are correlated; Extended Data Figure~\ref{fig:scinttime}), while the Amplitude Modulated Polarized Shot Noise frequency structure will change depending on the degree of polarization \citep{cordes_2004_apj}. SAMPSN predicts a correlation coefficient of \begin{equation}
    \rho = \frac{1}{2 + p_{\rm frac}^2},
\end{equation} 
where $p_{\rm frac}$ is the total polarization fraction. Our measured $\rho = 0.31$ implies that the signal is 100\,\% polarized, consistent with the high linear polarization fraction we measure in the frequency-averaged burst profiles (Figure~\ref{fig:bursts}), and the on-burst S/N distribution.

For B2 and B4, we find a weighted average correlation coefficient of the individual 1\,\us\ time bin spectra (above a S/N of 9) of 0.19\,$\pm$\,0.001 and 0.16\,$\pm$\,0.001, respectively, significantly lower than the $0.33$ expectation for SAMPSN \citep{cordes_2004_apj}, and the $0.31$ measured from B3, but also significantly greater than $0$. This is consistent with highly polarized SAMPSN (similar to B3), where potentially the lower correlation coefficient can be attributed to low S/N and/or sparseness of shot pulses. This is consistent with the lack of evidence for resolved shots of emission in the high time resolution profiles of bursts B2 and B4 (Extended Data Figure\,\ref{fig:isol}).

\subsection*{Resolved sub-microsecond emission?}

To test whether the sub-microsecond temporal structures we observe are isolated shots of emission or, alternatively, consistent with AMN, we compare the high S/N features with the {\it local} brightness distribution. For burst B3, which exhibits envelope fluctuations on the $\upmu$s level (Extended Data Figure\,\ref{fig:corrcoef}b), we define the local distribution as $\pm 1.5625\,\upmu$s ($\pm50$\,bins) around the brightest feature (which we wish to measure the significance of). We take this range of bins to ensure that we are including enough samples to measure the local distribution, and to ensure we are not taking too many samples such that we lose the information about the burst envelope {\it locally}. Using an Anderson-Darling test \citep{stephens_1974_asa}, we confirm that this local distribution is exponentially distributed. Note that since we cannot distinguish (by eye) between temporal spikes that are due to AMN versus individual emission spikes, we include all time bins (within the range mentioned), excluding the central feature, in the distribution; therefore, the probabilities measured are lower-limits. We fit a $\chi^2$-distribution with 2 degrees of freedom, using a least-squares fit, to the S/N values within the range defining the local distribution. This distribution is expected for 100\,\% polarized SAMPSN \citep{cordes_1976_apj}. For the high S/N, 2-bin wide feature in burst B3 (Time$=0$, Figure\,\ref{fig:31.25ns}), the probability of drawing this from the local distribution is p\,$=4\times 10^{-8} \times 100\,{\rm bins}/2 = 2\times 10^{-6} $ (Extended Data Figure\,\ref{fig:isol}), i.e. inconsistent with AMN, supporting that this component is a resolved, isolated shot of emission.

We repeated the same analysis on bursts B2 and B4, since they have sufficient S/N to study at the highest time resolution. For B2 we created the temporal profile at 31.25\,ns using the subband 1302--1318\,MHz, and for B4 we use subband 1398--1414\,MHz: in both cases this is the subband containing the brightest spectral feature (Extended Data Figure\,\ref{fig:isol}). In the case of B3, at the highest time resolution there are temporal fluctuations that are multiple bins wide. In contrast, the only bright structure in both B2 and B4 at the highest time resolution are unresolved single-bin spikes. Given that the frequency extent of the bright spectral feature is less than the subband width of 16\,MHz (we attribute the spectral features to scintillation; see below), this results in an effective time resolution lower than the sampling resolution. Single bin spikes in this case are therefore more likely to be consistent with the noise process. We tested the significance of the brightest unresolved structure using the same method as described for B3. Again, we confirm that the local distribution around the structure of interest is exponentially distributed using an Anderson-Darling test \citep{stephens_1974_asa}. The probability of the highest unresolved spike in B2 and B4 is p $=1\times10^{-4} \times 100\,{\rm bins} = 0.01$ and p\,$=4 \times10^{-4} \times 100\,{\rm bins} = 0.04$, respectively. Given the effective resolution argument above, combined with the high probabilities of occurring by chance, there is not strong evidence supporting that these structures are isolated, unresolved shots of emission. Instead, we conclude that these features are consistent with the $\chi^2$ AMN distribution.

\subsection*{Temporal ACF and power spectrum}
In Extended Data Figures~\ref{fig:acf_b3}--\ref{fig:acf_b4}, we present each coherently dedispersed subband of burst B3, B2 and B4, respectively, (covering the frequency extent of the burst) at 31.25\,ns resolution. This is essentially the burst dynamic spectrum at poor frequency resolution. We select the four brightest subbands that contain the most burst structure, and computed the ACF of the time profile. These ACFs were averaged together and shown in panel \textbf{e} of Extended Data Figures~\ref{fig:acf_b3}--\ref{fig:acf_b4}. The reason for computing the ACF in this manner, as opposed to creating a frequency-averaged profile and computing the ACF, is to limit the smearing due to inaccuracies in the incoherent dedispersion (the subbands are shifted by an integer number of bins). For all three bursts, we measure a characteristic timescale on the order of 10\,$\upmu$s, determined by fitting a Lorentzian function to the ACF (out to a time lag that is chosen by eye, to help the fitting distinguish between multiple timescales). This timescale is consistent with the full burst extent in time. In burst B3, we measure an additional two timescales: a clear 1.11\,\us\ timescale, and even a shorter timescale (40\,ns) consistent with temporal structure on the few-bin level (Extended Data Figure\,\ref{fig:acf_b3}f--h). For B4, we additionally measure a short-temporal scale suggesting that there is temporal structure of a few-bins in the B4 31.25\,ns profile (Extended Data Figure\,\ref{fig:acf_b4}g). Although, it is worth noting that the height of this narrow ACF feature relative to the wider ACF feature (height of the cyan Lorentzian relative to the green Lorentzian in Extended Data Figures\,\ref{fig:acf_b3}h and \ref{fig:acf_b4}g) is smaller for B4 than B3, implying either that the S/N of these temporal fluctuations are lower, or that there are fewer temporal features on this timescale. In B2, we see no evidence for power on shorter timescales in the ACF (Extended Data Figure~\ref{fig:acf_b2}e--f).

The burst power spectra are presented in panel \textbf{c} of Extended Data Figures~\ref{fig:acf_b3}--\ref{fig:acf_b4}. We perform a Bayesian maximum likelihood fit of a power law (red noise) plus constant (white noise) of the form
\begin{equation}
    f(\nu) = A\nu^{-\alpha} + C,
\end{equation}
where A is the amplitude, $\alpha$ is the slope of the power law, and $C$ is a white noise component, to the power spectrum, using the Stingray modelling interface \citep{huppenkothen_2019_apj}. Additionally, we fit a red noise plus white noise plus Lorentzian model to the data, to search for the presence of a quasi-periodic oscillation (QPO). This search was motivated by the hint of structure in the ACF of bursts from \rthree\ \citep{nimmo_2021_natas}, and additionally, the low amplitude, wide frequency bump seen by eye in the downsampled power spectrum of burst B3 (Extended Data Figure\,\ref{fig:acf_b3}c). First, we compute the Bayesian Information Criterion for each model fit, defined as 
\begin{equation}
    {\rm BIC} = -2{\rm ln}(L) + k{\rm ln}(n),
\end{equation}
where $L$ is the maximum likelihood of the fit, $k$ is the number of parameters in the model, and $n$ is the number of data points. A lower BIC implies the data is better represented by that model, although it does not mean that the preferred model is a good fit to the data. Since the number of model parameters $k$ is included in the BIC, models with more parameters are penalized to avoid overfitting the data. Here we computed the $\Delta$\,BIC = BIC$_{PL}-$BIC$_{PL+lor}$, where BIC$_{PL}$ is the BIC of the red noise model, and BIC$_{PL+lor}$ is the BIC of the red noise plus QPO model. For B3, we measure a $\Delta$\,BIC\,$=-12.1$, which is significantly in favor of the red noise model. We then measure the goodness-of-fit p-value of the red noise model, which is the fraction of 100 simulations (using the MCMC package {\tt emcee} \citep{foremanmackey_2013_pasp}) with a maximum likelihood lower than the likelihood of our fit. This p-value is 0.53, implying the fit to the data is good. Finally, we test the significance of the highest outlier in the residuals of our best-fit model. The residuals are defined as 
\begin{equation}
    R(\nu) = \frac{2P(\nu)}{M(\nu)}, 
\end{equation}
for the power spectrum $P(\nu)$ and best-fit model $M(\nu)$. By simulating 100 power spectra from the posterior distribution (using {\tt emcee}) and computing the highest outlier for each simulation, we can compute a p-value (the fraction of simulated outliers that are higher than our measured outlier). We find no significant outliers in the residuals, implying we have no evidence of a QPO in the data. The power spectra of B2 and B4 are also found to be consistent with red noise. A summary of the $\Delta$\,BIC, red noise goodness-of-fit p-value with power law index, and p-value of outliers are presented in Extended Data Table\,\ref{tab:ACF_ps}.

\subsection*{Fluence and luminosity}
The burst profiles, in S/N units, are converted to physical units (flux density, Jy) using the radiometer equation \citep{cordes_2003_apj}, using typical values for Effelsberg's 1.4\,GHz receiver temperature (20\,K) and gain (1.54\,Jy\,K$^{-1}$). We expect these system values to be accurate to within 20\%, which dominates the errors on the peak flux density and fluence (note that the peak flux density depends strongly on the time resolution used, and the values reported in Table\,\ref{tab:burst_properties} are calculated using a time resolution of 8\,\us). We additionally consider a sky background temperature of 0.8\,K, by extrapolating from the 408\,MHz map \citep{remazeilles_2015_mnras}, using a spectral index of $-$2.7 \citep{reich_1988_aa}, and adding a 3\,K contribution from the cosmic microwave background \citep{mather_1994_apj}. For the 8\,\us\ burst profiles, we report the peak S/N, peak flux density and fluence (measured in the $\pm$\,2$\sigma$ width region) in Table~\ref{tab:burst_properties}. We also report the isotropic-equivalent spectral luminosity of the bursts, taking the distance to \frb\ as 3.63\,$\pm$\,0.34\,Mpc (Kirsten et al. submitted). 

\subsection*{Burst temporal extent and spectral structure}
We performed a 2-dimensional Gaussian fit to the burst dynamic spectra (Figure\,\ref{fig:bursts}), to determine the burst extents in time and frequency. The measured widths were clearly underestimated compared with what can be seen by eye in the dynamic spectra, likely since the bursts are not well-modelled by a 2-dimensional Gaussian function. In the case of B1, B3, B4, and B5 we use the Gaussian mean in time as the Time$=$0 reference in Figure~\ref{fig:bursts}, and use this to calculate the burst time of arrival corrected to the Solar System Barycenter at infinite frequency, reported in Table~\ref{tab:burst_properties}. For B2, since there are two clear burst components, we fit a 2-dimensional Gaussian to each and determined the Time$=$0 reference as the center of the means of those Gaussians. Note that we do not fit pulse broadening functions to estimate the scattering timescale since the estimated scattering timescale from the Milky Way ISM is $\sim$\,50\,ns ($\ll 8$\,\us) at 1.4\,GHz \citep{cordes_2002_arxiv}, and is consistent with the frequency structure we attribute to scintillation (see below). 

To more accurately determine the burst widths, we performed a 2-dimensional autocorrelation of the dynamic spectra, and fit these with 2-dimensional Gaussian functions (Extended Data Figure~\ref{fig:2dacf}). Note the zero-lag noise spike is removed from this ACF. We convert the standard deviation of this Gaussian fit (in both time and frequency) to a full-width at half-maximum (FWHM) by multiplying the standard deviation by the factor $2\sqrt{2\mathrm{ln}(2)}$. We report the burst time width (t$_{\text{wid}}=1/\sqrt{2}\ \times$ FWHM) and frequency extent ($\nu_{\text{wid}}=$ FWHM) in Table~\ref{tab:burst_properties}. In Figure~\ref{fig:bursts}, the colored bars below each burst profile indicate the $\pm$\,2$\sigma$ width used for calculations of the fluence and polarization fractions. 

In addition to the frequency extent measured with the 2-dimensional ACF, there is another, narrower frequency scale evident in the one-dimensional frequency ACF (or simply by eye in the dynamic spectra of the bright bursts). Shown in Extended Data Figure~\ref{fig:scintbw}a is the frequency ACF after subtracting the larger frequency scale Gaussian. We perform a least-squares fit of a Lorentzian function to the center of this ACF (defined by eye using the clearly visible central feature in the ACF; Extended Data Figure~\ref{fig:scintbw}a). A Lorentzian frequency ACF is expected for scintillation \citep{rickett_1990_araa}. The fit function is of the form
\begin{equation}
    \frac{a}{x^2+\nu_{\text{scale}}^2} + b,
\end{equation}
where $a$ is the amplitude, $b$ is a vertical offset, and $\nu_{\text{scale}}$ is the scintillation bandwidth (defined as the half-width at half-maximum of the ACF \citep{rickett_1990_araa}). The scintillation bandwidth measurements are reported in Table~\ref{tab:burst_properties}. The Galactic ISM is expected to introduce a broadening due to scattering of $\sim50$\,ns (at 1.4\,GHz) along this line-of-sight \citep{cordes_2002_arxiv}, in rough agreement with our measured scintillation bandwidth ($1/(2\pi\Delta\nu_{\rm scint})\sim 27$\,ns). We therefore attribute this narrower spectral structure to scintillation from the Milky Way ISM. This interpretation is supported by the stronger correlation in the spectrum of two bursts (B1 and B2) separated by 4.3 minutes, compared to the lack of correlation between bursts B3 and B4, which are separated by 2.5\,hrs (Extended Data Figure~\ref{fig:scinttime}). The expected scintillation time is $\sim10$ minutes (at 1.4\,GHz) at high Galactic latitudes, which is dependent on an effective velocity (the assumption for Galactic sources is 100\,km\,s$^{-1}$)\citep{cordes_1991_apj,cordes_2002_arxiv}. This effective velocity is likely smaller for an extragalactic source than for a Galactic pulsar in the same line of sight, since the effective velocity is usually dominated by the velocity of the pulsar. Assuming an effective velocity of $\sim 30$\,km\,s$^{-1}$, the expected scintillation time will be $\sim30$ minutes (at 1.4\,GHz), consistent with the decorrelation timescale constraints using \frb. 

\subsection*{Polarization calibration}
To study the polarization properties of the bursts, we must first calibrate the data. In these observations, we did not record a noise diode scan to use for delay calibration; instead, we use the test pulsar data, also taken to inspect the general data quality, to calibrate the polarimetric data. This analysis strategy has been used successfully in previous work to determine the polarimetric properties of radio bursts from the Galactic magnetar SGR~1935+2154, detected using voltage data of the Westerbork RT1 VLBI backend \citep{kirsten_2021_natas}, and also in a study of the repeating FRB~20180916B, using data from the Effelsberg telescope as part of an EVN campaign \citep{nimmo_2021_natas}. Since we detected bursts on separate epochs, we calibrate the data using the test pulsar scan closest in time to the bursts. B1 and B2 are $<\,1$\,hr from the PSR~B0355+54 calibration scan, and B3 and B4 are $\sim 4$\,hr and $<\,2$\,hr from the PSR~B0355+54 scan, respectively. We note that we could not recover the polarimetric properties of burst B5 likely due to the low S/N of the burst, and so the April 28 epoch is omitted from the remainder of this section.

The data could exhibit leakage between the polarization channels, which we assume only significantly affects Stokes~V (defined as V=LL$-$RR using the PSR/IEEE convention \citep{vanstraten_2010_pasa}). In the pulsar data, without calibrating the leakage, we reproduce the circular polarization fraction to within 3\% and 1\% of the published values \citep{gould_1998_mnras}, for the Feb 20 and Mar 7 epochs, respectively. We, therefore, apply no leakage calibration.

A delay between the polarization hands is more crucial to correct for since it can significantly impact our interpretation of the linear polarization fraction and RM. Using the known RM of PSR~B0355+54, 79\,rad\,m$^{-2}$ \citep{taylor_1993_apjs}, we performed a brute-force search for the delay between polarization hands, $D$, that maximises the linear polarization, by rotating the data using the factor 
\begin{equation}
    \mathrm{e}^{-2i\mathrm{RM}({c^2}/{\nu^2})}\mathrm{e}^{-2i\nu \pi D},
\end{equation}
where $c$ is the speed of light and $\nu$ is the frequency in Hz. We searched for delays between $-15$ and $15$\,ns, in steps of 0.01\,ns. Note that previous work with the VLBI recorder at Effelsberg showed an instrumental delay of 5.4\,ns \citep{nimmo_2021_natas}. The estimated delays are $-0.18$\,ns and $-4.11$\,ns for the Feb 20 and Mar 7 observations, respectively. 

After removing the effect of the estimated delays, we performed RM synthesis \citep{brentjens_2005_aa} on the burst data to estimate the RMs of the bursts. The Faraday spectra for each burst are shown in Extended Data Figure~\ref{fig:polplot}a. 

To refine the RM and delay measurements further, we performed a joint least squares fit of Stokes~Q and U spectra normalised by the linear polarization L$=\sqrt{\mathrm{Q}^2+\mathrm{U}^2}$, using the following equations
\begin{equation} Q/L = \cos(2(c^2\mathrm{RM}/\nu^2 + \nu \pi D + \phi)),\end{equation}
\begin{equation} U/L = \sin(2(c^2\mathrm{RM}/\nu^2 + \nu \pi D + \phi)),\end{equation}
where $\phi$ is a linear combination of the absolute angle of polarisation on the sky (referenced to infinite frequency) and the phase difference between the R and L polarisation channels. The estimates of the delay and RM are given as initial guesses to the fit, and we simultaneously fit the pulsar and the two bursts from the same observational epoch. We force the delay to be the same between the pulsar and FRB data, fix to the known RM of the pulsar (79\,rad\,m$^{-2}$), and initially we allow for a different RM per burst. The measured delays are $-0.22$\,$\pm$\,0.02\,ns and $-4.16$\,$\pm$\,0.03\,ns, for the Feb 20 and Mar 7 observations, respectively, and the measured RMs are presented in Table~\ref{tab:pol_properties}. The QU-fits are shown in Extended Data Figure~\ref{fig:polplot}. The measured RMs have large uncertainties due to covariances between the fit parameters, which is difficult to alleviate due to the low number of rotations (due to either RM or delay) across the burst frequency extent. We conclude that the bursts in this work have consistent RM values (B2's RM is slightly over $3\sigma$ from B3, but this apparent difference should be verified in future observations of this source using an independent delay calibrator), which is also in agreement with the previously reported RM ($-29.8$\,rad\,m$^{-2}$)\citep{bhardwaj_2021_apjl}. We note that neither this work nor \citet{bhardwaj_2021_apjl} correct for the ionospheric contribution to the RM, but this effect is likely to be $\lesssim 2$\,rad\,m$^{-2}$.  

The data for each burst were then corrected for the measured delay and RM. The polarization position angle (PPA) was calculated using the following equation:
\begin{equation}
    \mathrm{PPA} = 0.5\,\mathrm{arctan}\left(\frac{U}{Q}\right),
\end{equation}
and, to correct for the parallactic angle, this is rotated by
\begin{equation}
    \theta = 2 \tan^{-1}\left({\frac{\sin({\rm HA})\cos(\phi_{\rm lat})}{(\sin(\phi_{\rm lat})\cos(\delta)-\cos(\phi_{\rm lat})\sin(\delta)\cos({\rm HA})}}\right),
\end{equation}
where ${\rm HA}$ is the hour angle of the burst, $\phi_{\rm lat}$ is the latitude of the Effelsberg telescope, and $\delta$ is the declination of \frb. Due to the fact that our observations did not feature an independent polarization calibrator scan, there remains an uncalibrated absolute phase offset in the data.  Therefore, we cannot compare the PPAs between our two observational epochs, or with other PPA measurements of \frb. By allowing for individual RM values per burst, we find that B1 and B2 exhibit a $\Delta$PPA of 28.5$^\circ$, and the $\Delta$PPA between B3 and B4 is 21.7$^\circ$. The unbiased linear polarization is computed following \citet{everett_2001_apj}, where
\begin{equation} \label{eq:Lunbias} L_{\rm unbias}=\begin{cases}
    \sigma_{I}\sqrt{\left(\frac{L_{\rm meas}}{\sigma_{I}}\right)^2-1}, & {\text{if}~  \frac{L_{\rm meas}}{\sigma_{I}}\ge 1.57} \\
    0, & \text{otherwise}
  \end{cases}  \end{equation} 
where $L_{\rm meas} = \sqrt{Q^2+U^2}$, and $\sigma_{I}$ is the standard deviation in the off-burst Stokes~I data. 

Bursts B1 -- B4 are highly linearly polarized ($>90$\%), and exhibit little-to-no evidence for circular polarization. There is a tentative 3 -- 4\,$\sigma$ detection of 13\,\% and 6\,\% circular polarization in B2 and B4, respectively. The linear and circular fractions, as well as the PPA offset from a weighted mean PPA of bursts from a single epoch, are presented in Table~\ref{tab:pol_properties}, and the polarization profile and PPA for each burst is shown in Figure~\ref{fig:bursts}.

It has been seen in the literature that some repeating FRBs exhibit a constant PPA ($\mathrm{\Delta(PPA) <\,10^\circ}$) and RM per observing epoch \citep{michilli_2018_natur,nimmo_2021_natas}. Therefore, in addition to the individual burst RM measurements, we fit for a global RM per observing epoch. In this global fit, we find that the PPA of B1 and B2 differ by 68.3$^\circ$ (Table~\ref{tab:pol_properties}). This large $\Delta$PPA may result from the low S/N of burst B1. The $\Delta$PPA between B3 and B4 in this global fit is 23.6$^\circ$, comparable to the difference we measured in the independent-RM fit above, unsurprising given the very similar RM values in the independent fits. 

\section*{Acknowledgements}
We thank W. van Straten for help with {\tt digifil}.
The European VLBI Network is a joint facility of independent European, African, Asian, and North American radio astronomy institutes. Scientific results from data presented in this publication are derived from the following EVN project code: EK048.
A.B.P is a McGill Space Institute (MSI) Fellow and a Fonds de Recherche du Quebec - Nature et Technologies (FRQNT) postdoctoral fellow.
B.M. acknowledges support from the Spanish Ministerio de Econom\'ia y Competitividad (MINECO) under grant AYA2016-76012-C3-1-P and from the Spanish Ministerio de Ciencia e Innovaci\'on under grants PID2019-105510GB-C31 and CEX2019-000918-M of ICCUB (Unidad de Excelencia ``Mar\'ia de Maeztu'' 2020-2023).
C.L. was supported by the U.S. Department of Defense (DoD) through the National Defense Science \& Engineering Graduate Fellowship (NDSEG) Program.
D.M. is a Banting Fellow.
E.P. acknowledges funding from an NWO Veni Fellowship.
F.K. acknowledges support from the Swedish Research Council.
FRB research at UBC is supported by an NSERC Discovery Grant and by the Canadian Institute for Advanced Research.
J.P.Y. is supported by the National Program on Key Research and Development Project (2017YFA0402602).
K.S. is supported by the NSF Graduate Research Fellowship Program.
K.W.M. is supported by an NSF Grant (2008031).
M.B. is supported by an FRQNT Doctoral Research Award.
N.W. acknowledges support from the National Natural Science Foundation of China (Grant 12041304 and 11873080).
P.S. is a Dunlap Fellow and an NSERC Postdoctoral Fellow. The Dunlap Institute is funded through an endowment established by the David Dunlap family and the University of Toronto.
V.B. acknowledges support from the Engineering Research Institute Ventspils International Radio Astronomy Centre (VIRAC).
Work at UvA and ASTRON was funded by the NWO Vici grant ``AstroFlash" (PI: Hessels, VI.C.192.045).
%

%\bibliography{bib-entries}{}
%\bibliographystyle{aasjournal}
%\input{bib_bbl.bbl}

\paragraph{Author contributions}
K.N. led the data analysis, made the figures, and wrote most of the manuscript. J.W.T.H. guided the work and made important contributions to the writing and interpretation.  F.K. discovered the bursts and contributed to the analysis of the voltage data. A.K. adapted the SFXC code to created coherently dedispersed voltage data at the native time resolution.  J.M.C. provided important insights into the data analysis strategy.  M.P.S., D.M.H. and R.K. played supporting roles in the data acquisition and analysis.  All other authors contributed significantly to laying the groundwork for this study, aspects of the data acquisition, or interpretation.

\paragraph{Competing interests}
The authors declare no competing interests.

\paragraph{Additional information}
\paragraph{Correspondence and requests for materials} should be addressed to K.N.

\setcounter{figure}{0}
\captionsetup[figure]{name={\bf Extended Data Figure}}

\setcounter{table}{0}
\captionsetup[table]{name={\bf Extended Data Table}}

\begin{figure*}
\resizebox{\hsize}{!}
        {\includegraphics[trim=0cm 0cm 0cm 0cm, clip=true]{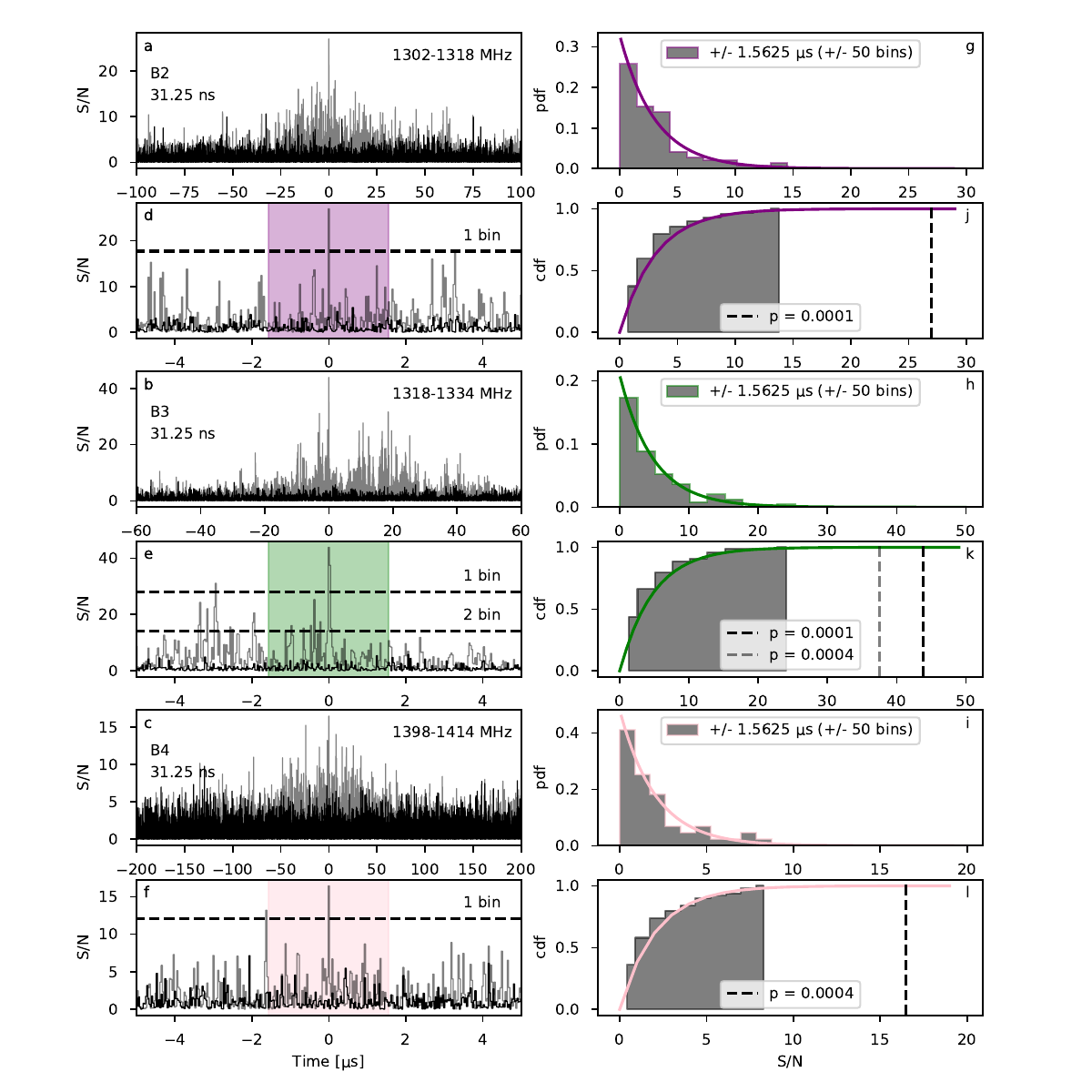}}
  \caption{The probability that the brightest 1--2 bin features in the 31.25\,ns burst profiles are consistent with the local amplitude modulated noise distribution. Panels \textbf{a}--\textbf{c} show the 31.25\,ns resolution profiles (grey) with off burst noise (black) shown for comparison. The burst name and time resolution is shown in the top left corner and the frequency range averaged over to produce the burst profile in the top right corner of the panels. Panels \textbf{d}--\textbf{f} are zoomed-in profiles containing the highest S/N feature in the burst profile. The colored region represent the local region (time span $\pm 1.5625$\,\us) used to determine the probability density function (pdf; panels \textbf{g}--\textbf{i}) and cumulative density function (cdf; panels \textbf{j}--\textbf{l}). Note that the feature at the center of the colored region is not added to the distribution, since this is the feature which we want to determine the significance of relative to the local distribution. An exponential distribution fit is overplotted (colored lines) on the pdf and cdf. The highest S/N feature is represented by the vertical dashed line on the cdf (where in the case of B3, there are two dashed lines since the feature is 2 bins wide), and the legend shows the probability (or 1-cdf) of these features. The horizontal dashed lines on panels \textbf{d}--\textbf{f} represent the 3\,$\sigma$ levels for single-bin features, using this local distribution (also for 2-bin features in the case of B3).}
     \label{fig:isol}
\end{figure*}

\begin{figure*}
\resizebox{\hsize}{!}
        {\includegraphics[trim=0cm 0cm 0cm 0cm, clip=true,width=\textwidth,height=195mm]{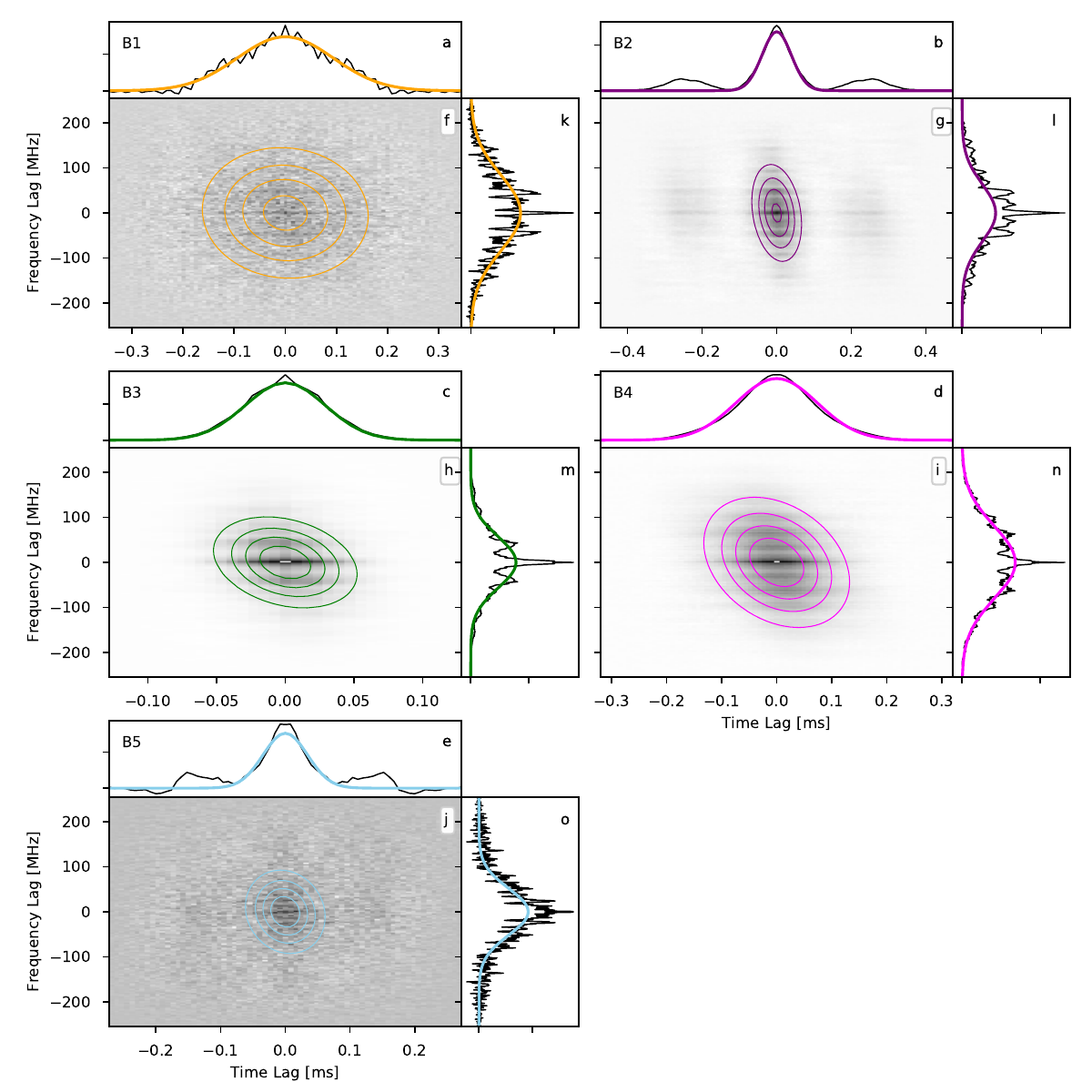}}
  \caption{Low time resolution 2D autocorrelation functions (ACFs) of the bursts detected from \frb. Panels \textbf{f}--\textbf{j} show the 2D ACF with colored contours overplotted representing the 2D Gaussian fit 1,2,3 and 4 $\sigma$. The zero lag spike is not plotted. The ACF is computed using filterbank data generated with SFXC \citep{keimpema_2015_exa}, with time and frequency resolution of 8\,$\upmu$s and 125\,kHz, respectively. The data were dedispersed using a DM of 87.75\,\dmunit. Panels \textbf{a}--\textbf{e} show the frequency-averaged time ACF, with the frequency-averaged Gaussian fit overplotted, and similarly panels \textbf{k}--\textbf{o} show the time-averaged frequency ACF, with the time-averaged Gaussian fit overplotted. The time and frequency scales characterised in this plot, arise from the burst temporal width and frequency extent. The colored lines coordinate with other figures in this work (e.g. Figure~\ref{fig:bursts} and Extended Data Figure~\ref{fig:polplot}).}
     \label{fig:2dacf}
\end{figure*}

\begin{figure*}
\resizebox{\hsize}{!}
        {\includegraphics[trim=0cm 0cm 0cm 0cm, clip=true]{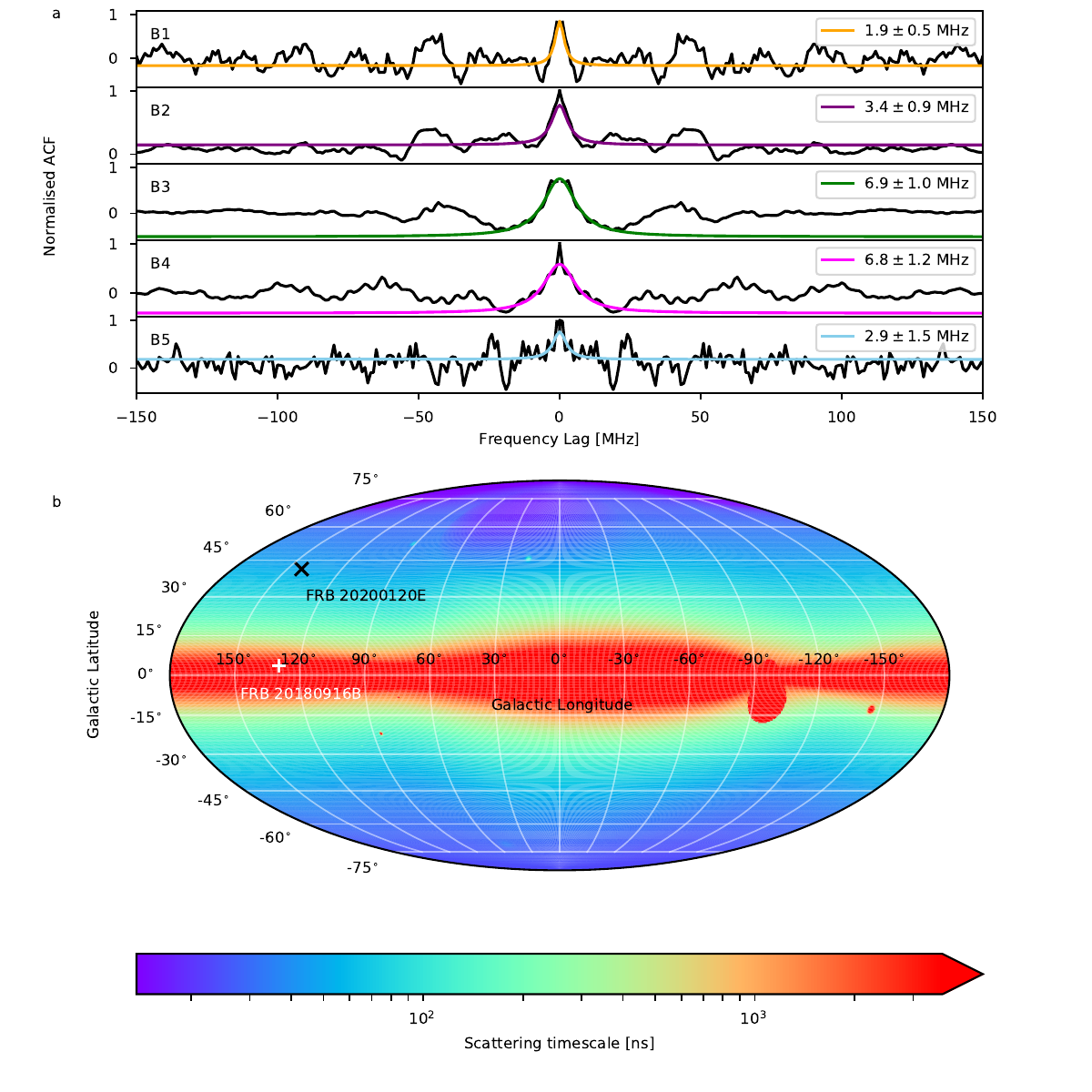}}
  \caption{Measurement of the scintillation bandwidth in the autocorrelation function (ACF) of each of the five bursts from \frb. Sub-figure \textbf{a} shows the time-averaged frequency ACF from Extended Data Figure~\ref{fig:2dacf} after subtracting the Gaussian fit (black). The Lorentzian fit to the central component is shown by the colored line. The burst name is shown in the top left of each panel, and the measured scintillation bandwidth (defined as the half-width at half-maximum of the Lorentzian; \citealt{rickett_1990_araa}) is shown in the top right of each panel. Sub-figure \textbf{b} shows a colormap of the expected scattering timescale at 1.4\,GHz as a function of Galactic longitude and latitude, from the NE2001 Galactic electron density model \citep{cordes_2002_arxiv}. The sky positions of both FRB~20200120E and FRB~20180916B are shown by the black cross and white plus, respectively.   }
     \label{fig:scintbw}
\end{figure*}

\begin{figure*}
\resizebox{\hsize}{!}
        {\includegraphics[trim=0cm 0cm 0cm 0cm, clip=true]{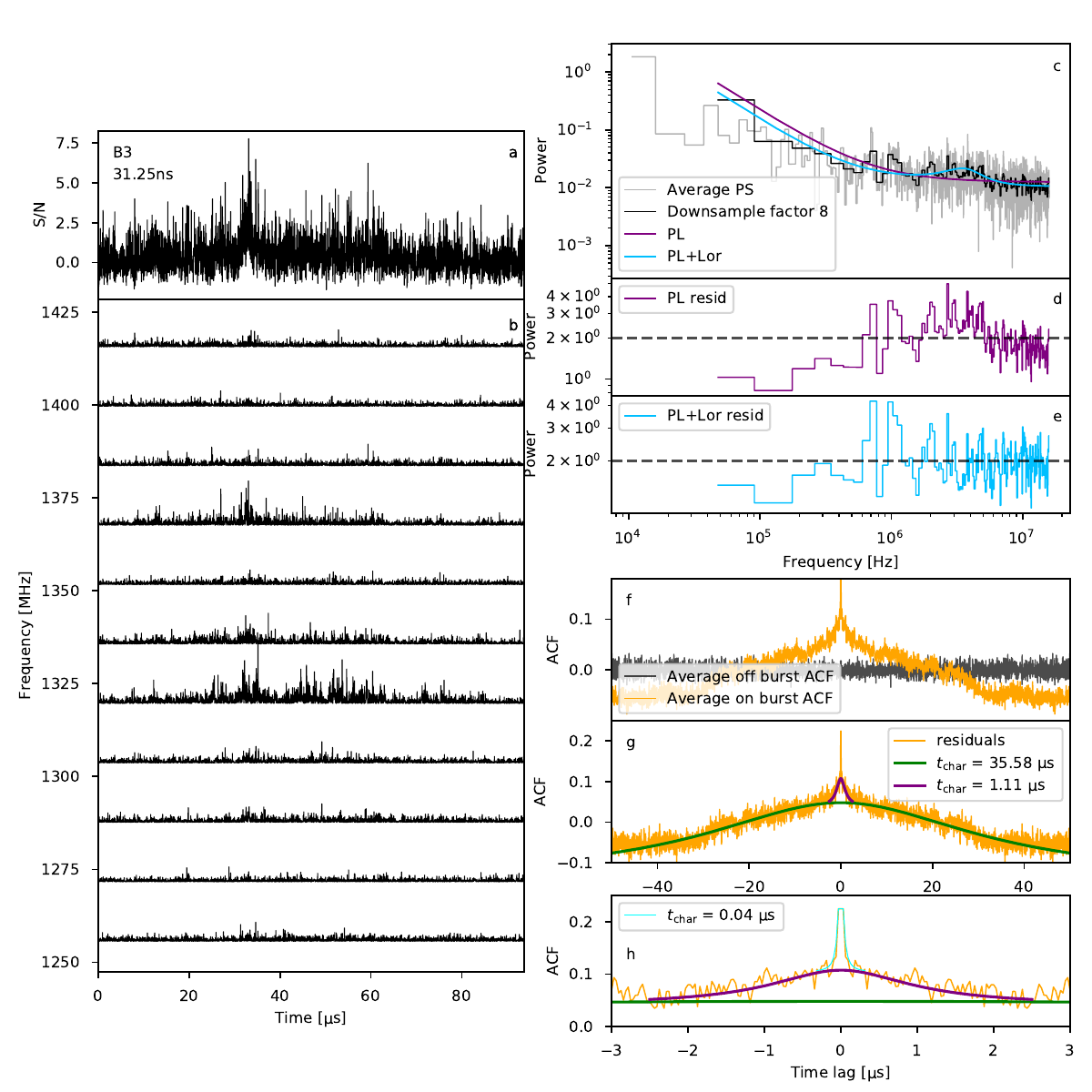}}
  \caption{Dynamic spectrum of burst B3 from \frb\ with time resolution 31.25\,ns, temporal autocorrelation function (ACF) and power spectrum (PS). Panel \textbf{b} shows the dynamic spectrum in the form of temporal profiles per subband. This data was generated with SFXC, and each subband has been coherently dedispersed to 87.7527\,\dmunit. Panel \textbf{a} shows the frequency-averaged burst profile. Panel \textbf{c} shows the average power spectrum (PS) of the four subbands containing significant burst structure in the top panel (grey), with a downsampled PS (factor 8) overplotted in black. The purple and blue lines represent fits to the PS of a red noise power law plus white noise model and a power law/white noise plus Lorentzian model, respectively. Panels \textbf{d} and \textbf{e} below show the residuals ($2\times D/M$, for data $D$ and model $M$) of both models, matching the colors above. The dashed lines represent the perfect case of $D=M$. Panel \textbf{f} shows the average temporal ACF of the same four subbands (top panel, orange). For comparison the off burst ACF is also shown (grey). The residual of the average ACF subtracted the noise ACF is shown in panel \textbf{g} below, with the green and purple Lorentzian fits to the ACF residuals highlighting two distinct temporal scales in the data. Panel \textbf{h} shows a zoom in on the ACF residuals highlighting a third temporal scale by the cyan Lorentzian fit.}
     \label{fig:acf_b3}
\end{figure*}

\begin{figure*}
\resizebox{\hsize}{!}
        {\includegraphics[trim=0cm 0cm 0cm 0cm, clip=true]{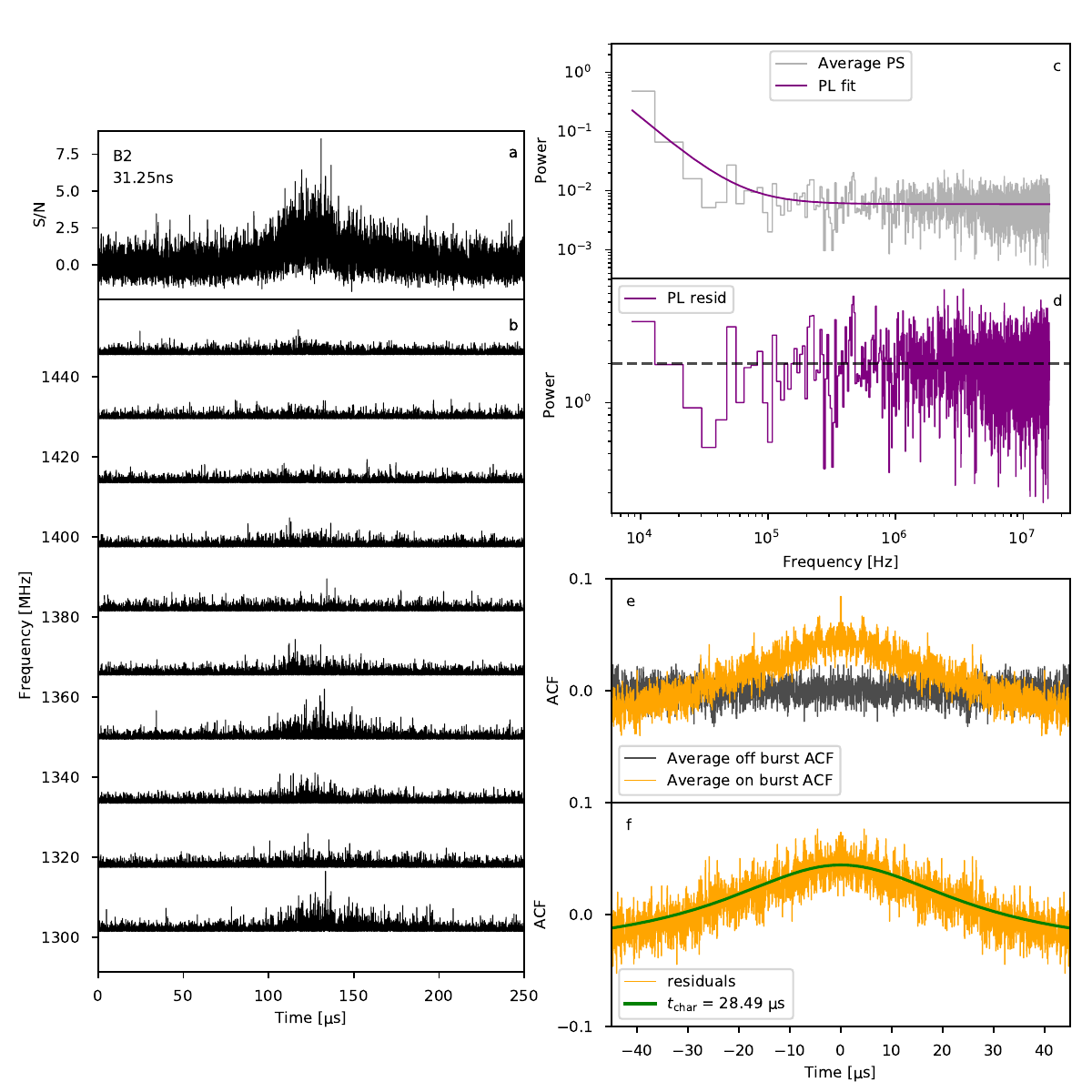}}
  \caption{The same as Extended Data Figure~\ref{fig:acf_b3}, for burst B2 from \frb. Note the average ACF only shows one temporal scale (unlike the three seen for burst B3). Additionally, we only plot the red noise power law plus white noise model since any wide Lorentzian features are less apparent in this power spectrum, than the case of B3.}
     \label{fig:acf_b2}
\end{figure*}

\begin{figure*}
\resizebox{\hsize}{!}
        {\includegraphics[trim=0cm 0cm 0cm 0cm, clip=true]{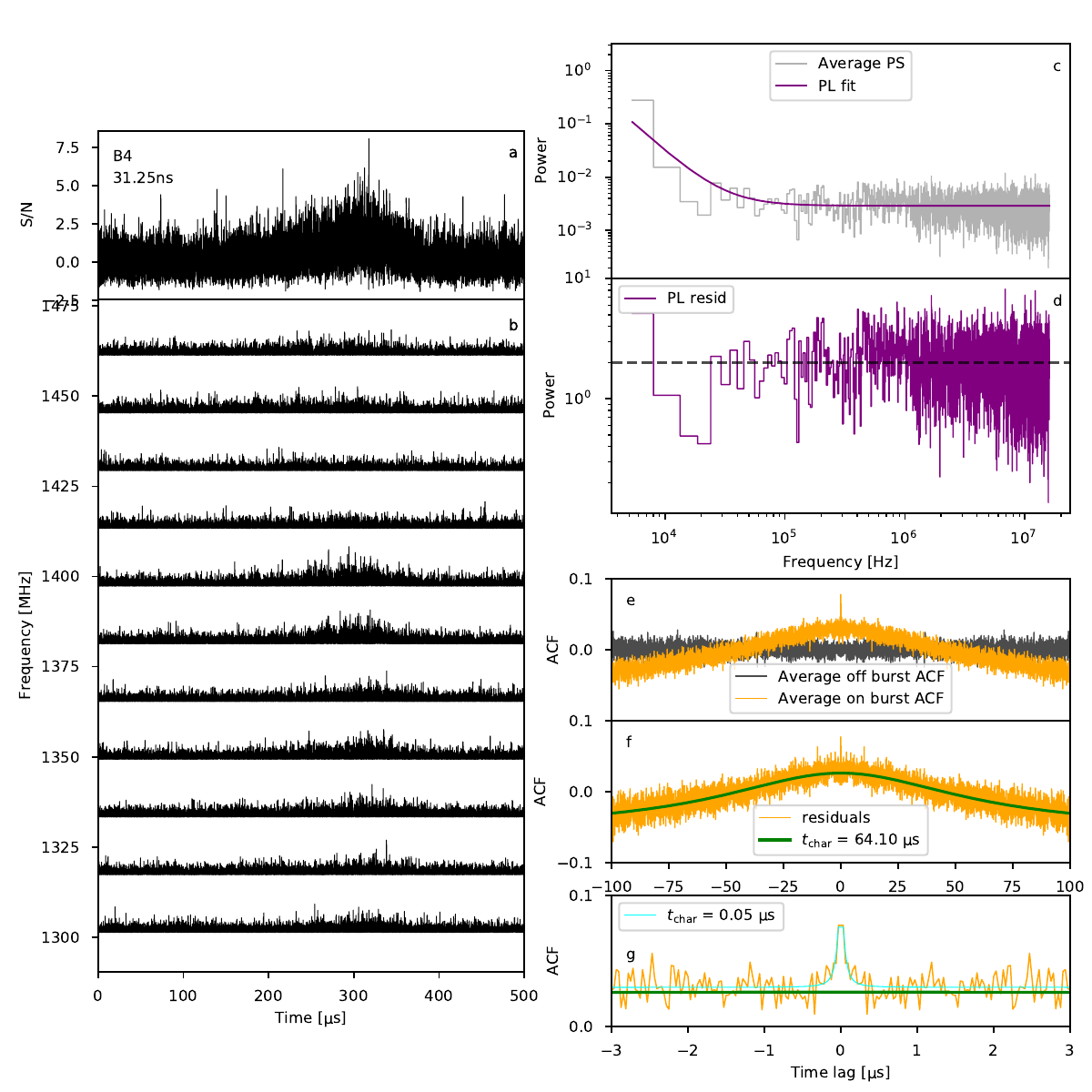}}
  \caption{The same as Extended Data Figures~\ref{fig:acf_b3} and \ref{fig:acf_b2}, for burst B4 from \frb. Note the average ACF shows two temporal scales (unlike the $3$ seen for burst B3). Additionally, we only plot the red noise power law plus white noise model since any wide Lorentzian features are less apparent in this power spectrum, than the case of B3.}
     \label{fig:acf_b4}
\end{figure*}

\begin{figure*}
\resizebox{\hsize}{!}
        {\includegraphics[trim=0cm 0cm 0cm 0cm, clip=true]{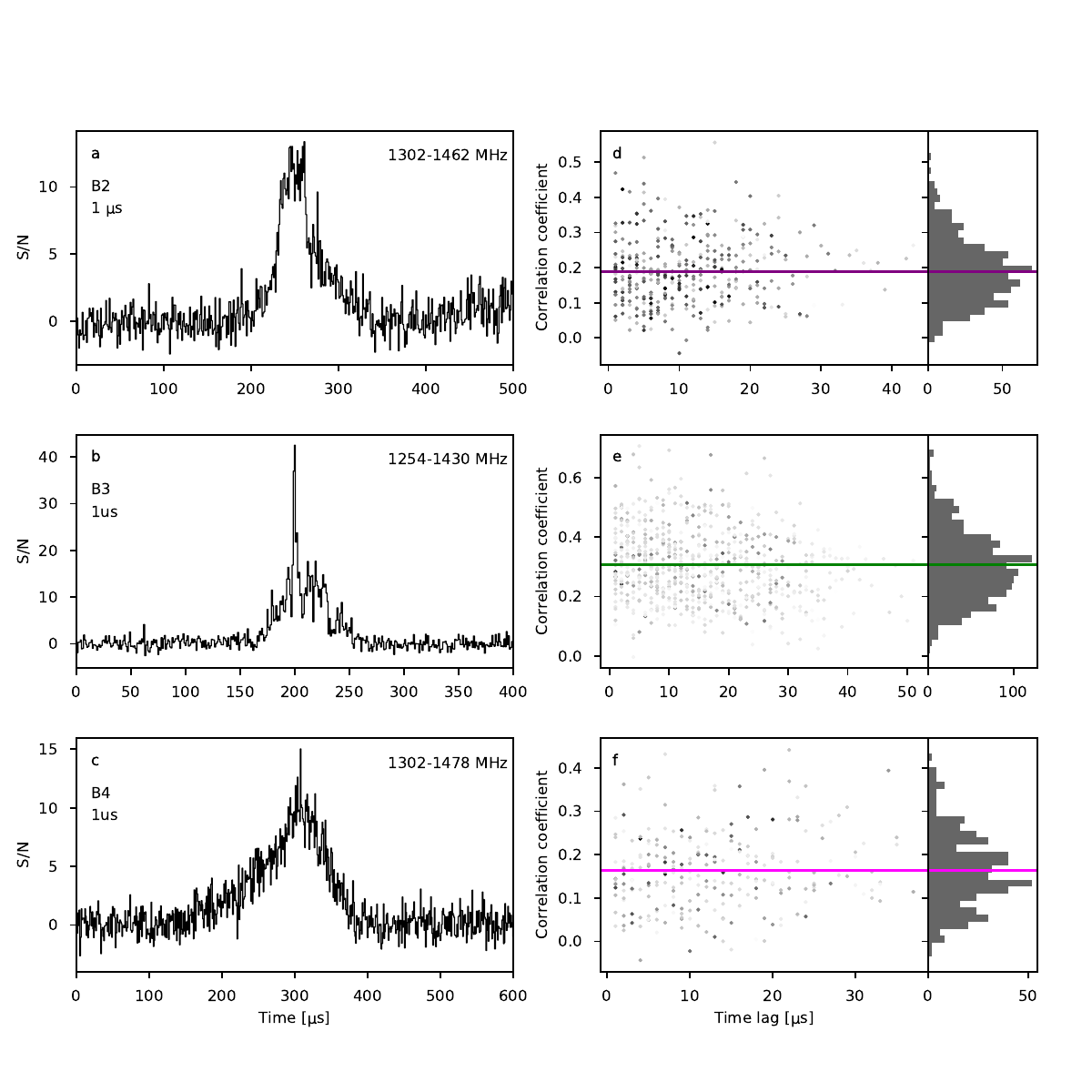}}
  \caption{Correlation coefficient between single time bin (1\,\us) spectra of bursts B2, B3 and B4 from \frb, testing for consistency with the Scintillating Amplitude Modulated Polarized Shot Noise prediction (0.33 for 100\,\% polarized emission \citep{cordes_2004_apj}). Panels \textbf{a}--\textbf{c} show the 1\,\us\ burst profile with the burst name and time resolution shown in the top left corner and the frequency range averaged over to produce the burst profile in the top right corner of each panel. Panels \textbf{d}--\textbf{e} show the correlation coefficient between single time bin spectra above a S/N threshold of $9$ as a function of the time separation between bins. The color gradient indicates the geometric mean of the two time bins used to determine the correlation coefficient (darker color implying a higher geometric mean). Also plotted is a histogram of the correlation coefficients. The colored line represents the geometric mean S/N weighted correlation coefficient. }
     \label{fig:corrcoef}
\end{figure*}

\begin{figure*}
\resizebox{\hsize}{!}
        {\includegraphics[trim=0cm 0cm 0cm 0cm, clip=true,width=\textwidth,height=195mm]{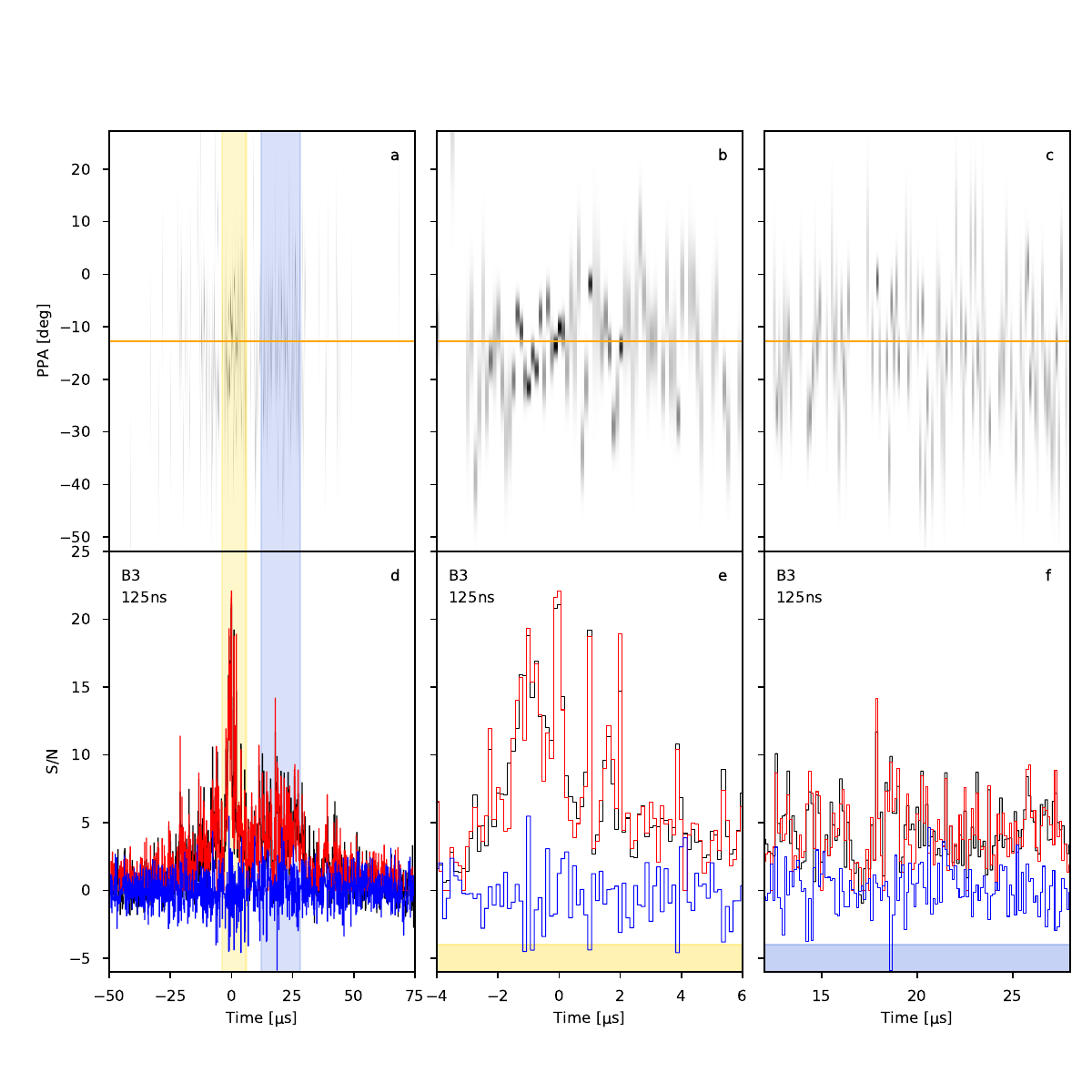}}
  \caption{High time resolution polarimetric profile and polarization position angle (PPA) for burst B3 from \frb. Panels \textbf{a}--\textbf{c} show the PPA as a function of time, with the orange line representing the weighted best-fit line to the PPA. Only the PPAs above a linear S/N threshold of $5$ are plotted.  Panels \textbf{d}--\textbf{f} show the polarimetric profile of the burst sampled at 125\,ns, with Stokes~I (black), unbiased linear polarization (\citealt{everett_2001_apj}; red) and circular polarization (blue). The yellow and blue regions plotted on panels \textbf{a} and \textbf{d} represent the time ranges used for plotting panels \textbf{b},\textbf{e} and panels \textbf{c},\textbf{f}, respectively. This data was generated with SFXC, with 4\,MHz channels and coherently (within subbands) and incoherently (between subbands) dedispersed to 87.7527\,\dmunit. The frequency information was averaged for the frequency range 1254 -- 1430\,MHz (visually, the extent of the burst in frequency), which, in this data product, corresponds to averaging by a factor of 44.}
     \label{fig:B3_125ns_fullpol}
\end{figure*}

\begin{figure*}
\resizebox{\hsize}{!}
        {\includegraphics[trim=0cm 0cm 0cm 1cm, clip=true]{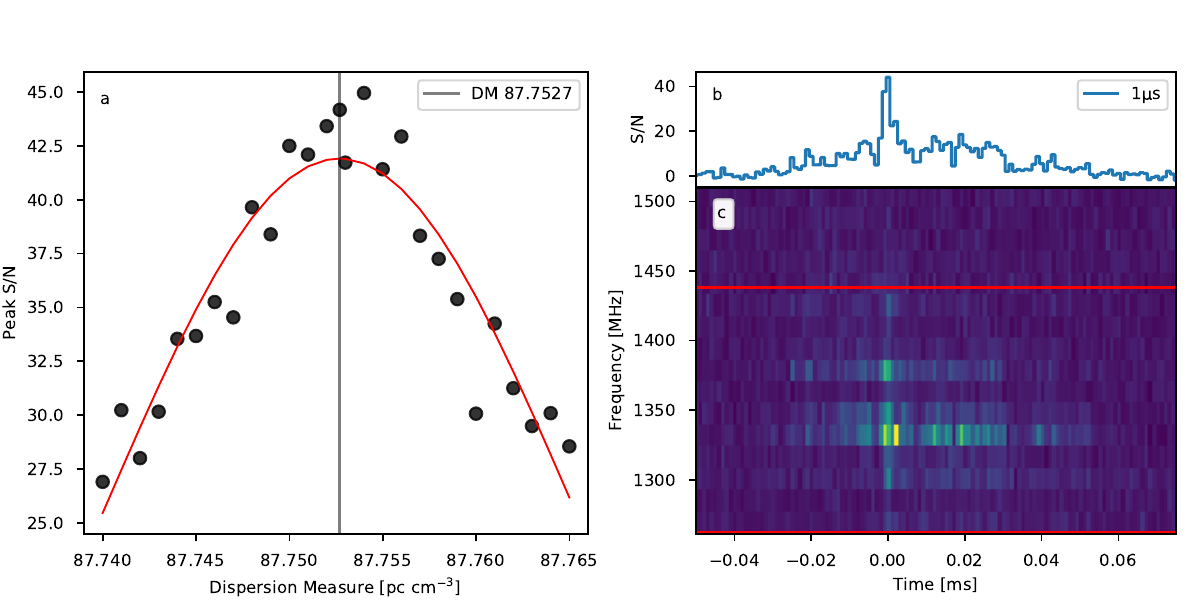}}
  \caption{Constraining the dispersion measure (DM) using the short timescale structure in burst B3 from \frb. Using data products generated using SFXC \citep{keimpema_2015_exa}, with a time and frequency resolution of 500\,ns and 1\,MHz, respectively, we coherently (within subbands) and incoherently (between subbands) dedisperse to a range of DMs, downsample in time by a factor of 2, and compute the peak S/N of the frequency-averaged profile (panel \textbf{a}). Also plotted in panel \textbf{a} is a Gaussian fit to the peak S/N as a function of DM, with the best-fit DM (DM = 87.7527$\pm$0.0003\,\dmunit) shown by the grey line. Panel \textbf{b} shows the burst profile and panel \textbf{c} the dynamic spectrum coherently and incoherently dedispersed to the best-fit DM. The time and frequency resolutions used for plotting are 1\,\us\ and 16\,MHz, respectively. The red lines on the dynamic spectrum indicate the frequency extent averaged over to determine the peak S/N per DM, and to produce the burst profile.   
   }
     \label{fig:DM}
\end{figure*}

\begin{figure*}
\resizebox{\hsize}{!}
        {\includegraphics[trim=0cm 0cm 0cm 0cm, clip=true]{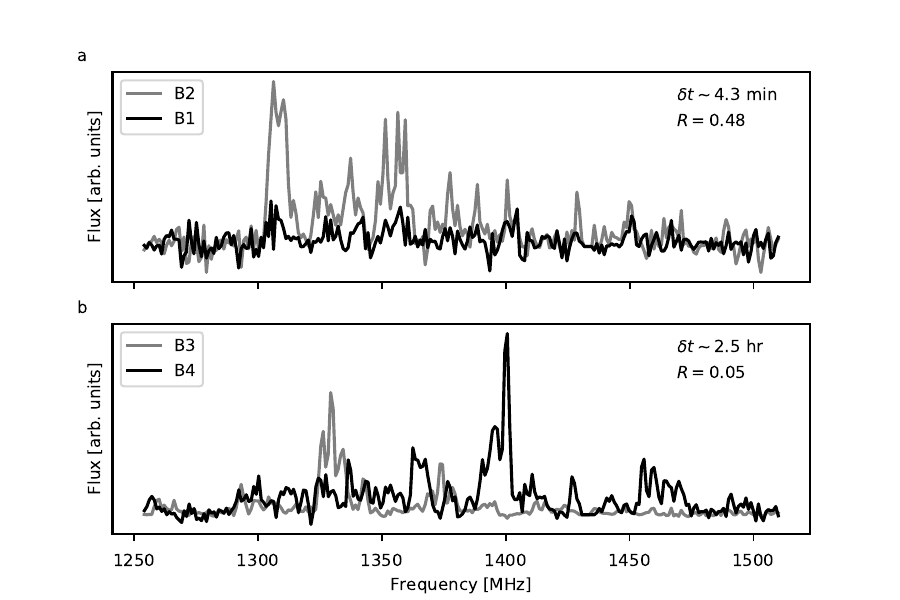}}
  \caption{Comparing the time-averaged spectra between bursts detected close in time. Panel \textbf{a} shows the time-averaged spectra for B1 (black) and B2 (grey). Panel \textbf{b} shows the time-averaged spectra for B3 (grey) and B4 (black). In the top right of each panel, we quote the time between the two bursts in the plot, $\delta t$, and the correlation coefficient of the burst spectra, $R$.}
     \label{fig:scinttime}
\end{figure*}

\begin{table*}
\caption{\label{tab:ACF_ps}High time resolution autocorrelation function and power spectra results.}
\resizebox{\textwidth}{!}{\begin{tabular}{lcccccc}
\hline
\hline
		{Burst} & {Characteristic timescales [$\upmu$s}] & {Red. $\chi^2$ $\mathrm{^{a}}$} & {Power law index $\mathrm{^{b}}$} & {$\Delta$ Bayesian Information Criterion$\mathrm{^{c}}$} & {Goodness of fit p-value$\mathrm{^{d}}$} & {Outlier p-value$\mathrm{^{e}}$}  \\
		\hline
		B2 & 28.5\,$\pm$\,0.2 & 2.0 & $1.85\,\pm\,0.04$ & $-$20.5 & 0.50 & 0.98 \\
		B3 & 0.04\,$\pm$\,0.3, 1.1\,$\pm$0.4, 35.6\,$\pm$0.5 &  2.4, 1.64, 1.3 & $1.46\,\pm\,0.05$ & $-12.1$ &  0.53 & 0.99 \\
		B4 & 0.05\,$\pm$\,0.02, 64.1\,$\pm$\,1.5 & 2.7,1.9 & $2.04\,\pm\,0.05$ & $-$15.4 & 0.66 & 0.97 \\
		\hline
	    \multicolumn{6}{l}{$\mathrm{^{a}}$ The reduced $\chi^2$ of the Lorenzian fit to the high time resolution ACF in Extended Data Figures~\ref{fig:acf_b3}--\ref{fig:acf_b4}.}\\
        \multicolumn{6}{l}{$\mathrm{^{b}}$ The power law index measured for the power spectra.}\\
        \multicolumn{6}{l}{$\mathrm{^{c}}$ Metric for model comparison (BIC for power law red noise model $-$ BIC for power law plus Lorentzian). }  \\
        \multicolumn{6}{l}{$\mathrm{^{d}}$ Goodness of fit of the best fit model.}  \\
    
    \multicolumn{6}{l}{$\mathrm{^{e}}$ p-value of the highest outlier in the residuals of the power spectrum.}  \\
\end{tabular}}
\end{table*}

\begin{figure*}
\resizebox{\hsize}{!}
        {\includegraphics[trim=0cm 0cm 0cm 0cm, clip=true,width=\textwidth,height=195mm]{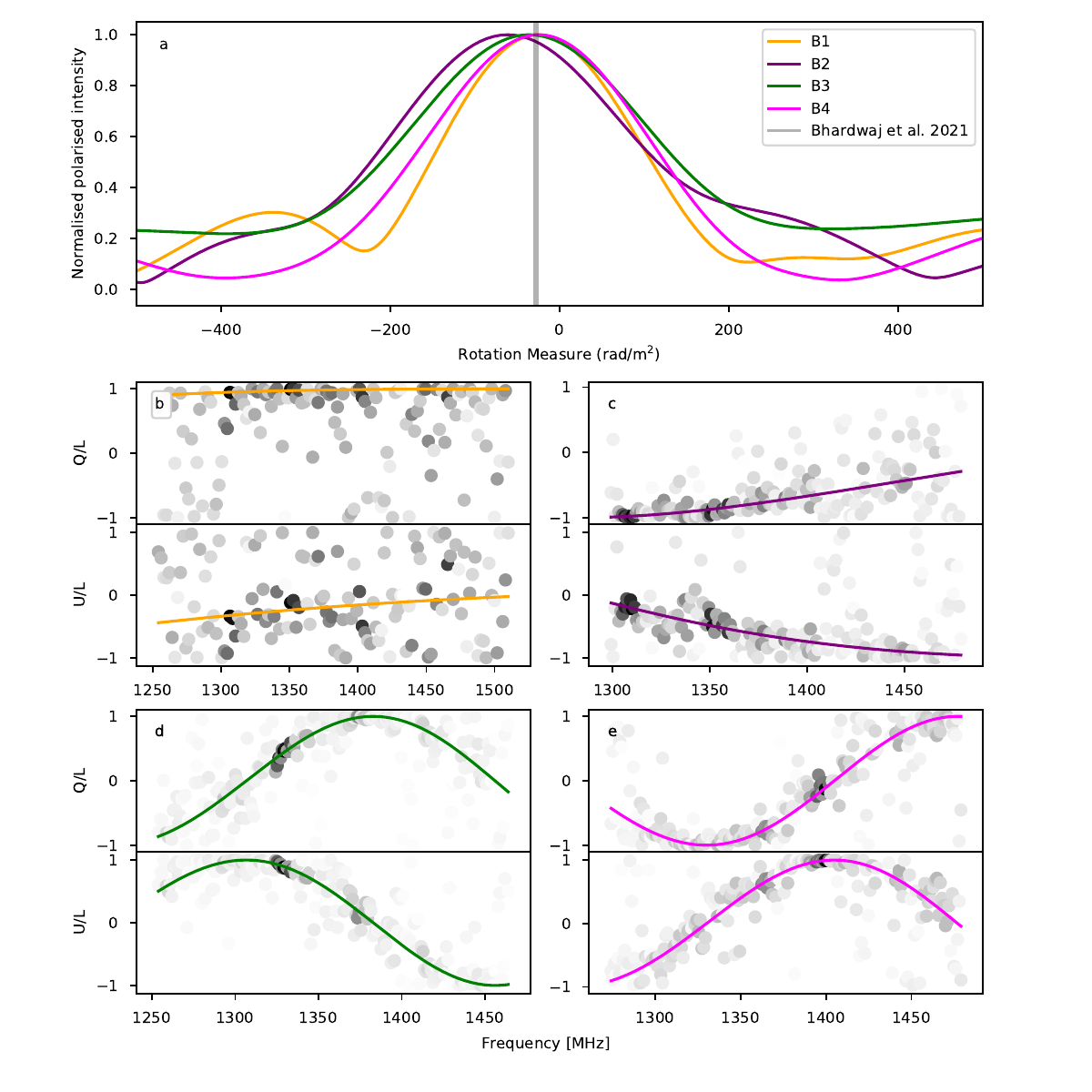}}
  \caption{Rotation measure (RM) determination for B1 -- B4 from \frb. Panel \textbf{a} shows the Faraday spectrum per burst (labelled), determined using RM synthesis \citep{brentjens_2005_aa}, with the grey line showing the previous RM measurement of this source \citep{bhardwaj_2021_apjl}. The grey-scale scatter points panels \textbf{b}--\textbf{e} show the Stokes~Q (top) and Stokes~U (bottom) spectra normalised by the linear polarization L$=\sqrt{\mathrm{Q}^2+\mathrm{U}^2}$, where the darker color represents a higher S/N. The colored lines show the best-fit QU-fitting result, where we fit for both the RM and the delay between the polarization channels (Methods). The color of the line indicates which burst is being plotted, using the legend in the top figure.}
     \label{fig:polplot}
\end{figure*}

\end{document}